\documentclass[aps,preprint,a4paper,showpacs,amsmath,amssymb,floatfix]{revtex4}

\usepackage{lscape,graphicx}
\usepackage{epsfig} 

\newcommand{\Op}[1]{{\bf {\hat {#1}}}}
\begin{document}

\include{epsf}

\title{  Grid methods for cold molecules : determination of photoassociation lineshapes and rate constants}
\author{S.\ Kallush$^{1,2}$, R. Kosloff$^{2}$ and F. Masnou-Seeuws$^{1}$}
\affiliation
{$^1$ Laboratoire Aim\'{e} Cotton, CNRS, B\^{a}t. 505, Campus d 'Orsay, 91405 Orsay Cedex, France
\\$^2$The Fritz Haber Research Center, The Hebrew University of Jerusalem,
Jerusalem 91904, Israel}

\begin{abstract}
A description of photoassociation by CW laser is formulated in the framework of grid methods. The Hamiltonian describing one or several bound states coupled to a multiple of continuum manifolds via a radiative field is written in the energy representation and diagonalized. The generality of the treatment 
allows to compute accurately and efficiently physical properties such as non-linear high-intensity energy shifts, line shapes, and photoassociation rates both for isolated and non-isolated resonances. Application is given to sodium photoassociation in the experimental conditions of Mc Kenzie {\it et al} [Phys. Rev. Lett. {\bf 88}, 090403 (2002)]. Inverted region for the dependency of the rate vs. the intensity and non-symmetric lineshapes were predicted to occur above the saturation limit. Comparison with the  model of Bohn and Julienne [Phys. Rev. A {\bf 60}, 414 (1999)] is discussed.

\end{abstract}


\maketitle

\section{Introduction}

Many experimental groups are presently working on the formation of ultracold molecules and molecular quantum gases. A very efficient scheme is the photoassociation reaction, where two colliding atoms absorb a photon to make a molecule \cite{thorsheim87}; moreover ultracold photoassociation spectroscopy \cite{jones2006} is capable of providing deep insight into the properties of long range molecules, cold atomic collisions, or pair correlations in a condensate. Theoretical two-body scattering models have been developed to compute the photoassociation rates \cite{pillet97,napolitano97,cote98,bohn99,taylor2004}, in the framework of a perturbative treatment, adapted to situations where low-intensity CW lasers are employed and many-body effects are negligible.  A linear variation of the photoassociation rate as a function of intensity is then predicted.  In the work of Bohn and Julienne \cite{bohn99}, which develops a semi-analytic theory yielding laser-induced energy shifts and l
 ine broadening, the  departure from this linear behavior is described,  the saturation limit  at larger laser intensities being  evaluated. In a condensate, the two-body model should fail and  Javanainen and Mackie \cite{javanainen2002} have developed  a many-body theory which predicts  a saturation limit  occurring at lower intensities and attributed to rogue photodissociation. The search for saturation effects has stimulated several experiments. Measurement, by  Mc Kenzie {\it{et al }} \cite{mckenzie2002}, of the photoassociation rate in a sodium condensate  as a function of laser intensity has shown a linear behavior in the intensity range considered, in agreement with the two-body Bohn and Julienne estimation, and in contrast with the prediction of Javanainen and Mackie. For a quantum degenerate lithium atomic gas, measurement \cite{prodan2003} of the intensity dependence of the photoassociation rate has verified both the large value of this rate (at a temperature $T$=600 nK) and the two-body saturation effect at intensities $I \ge$ 30 W/cm$^2$. The agreement with theory is considered as reasonable within the present experimental accuracy: the higher value of the measured  saturation limit, and the oscillations in the experimental rate in the 30-80 W/cm$^2$ intensity range should be confirmed by measurements with a smaller error bar. For a non-degenerate gas of cesium atoms, at a temperature of 40 $\mu$K, within an optical dipole trap, saturation limit is reached for intensities above 100 W/cm$^2$\cite{kraft2005}, and the experimental values then lay around the theoretical curve : it is not clear whether such oscillations are due to uncertainties or should be attributed to a physical effect. \\

Recently Naidon and Masnou-Seeuws \cite{naidon2006} have developed a time-dependent many-body theory of photoassociation in an atomic Bose-Einstein condensate, showing that several physical effects depend on pair correlations. They show that the two existing  models for the description of such correlations  \cite{kohler2003b,naidon2003},  differ in  their predictions  of  the photoassociation line shapes at large intensities.  Precise measurements of the variation of the line shapes as a function of cw laser intensity should  therefore give insight into the optimal description of many-body effects. \\

The aim of the present paper is to anticipate such measurements by revisiting  the theoretical treatment of photoassociation  line shapes, taking advantage of the efficiency of grid methods \cite{kgrid96,fatal,slava99,willner2004}  for the numerical description of cold collisions and long range molecules, particularly in cases where several channels are involved. Actually, the later methods have been widely used for the computation of photoassociation rates \cite{dion01,kerman04,azzizi2004,sage2005}, since they can determine very precise energies and wave functions,  both for bound and for continuum levels. It has been demonstrated that due to their global character,  mapped grid  methods are ideally  suited to precise  computation of  the perturbations occurring in  molecular spectra, such as Rb$_2$\cite{slava00a,slava00b,bergeman2006} or {K$_2$}\cite{lisdat01} singlet-triplet mixing. In such cases, the perturbation of one level of a given series should be attributed to  a
  large number of bound or continuum levels of the other series, and local perturbation treatments are poorly adapted. However, up to know, attempts to compute the photoassociation line shapes via grid methods have not been very efficient \cite{pellegrini2003,naidon2004} and the interpretation of precise measurements is routinely performed with Numerov type approaches \cite{portier2006}. The present paper is proposing an alternative way of implementing grid methods for the calculation of photoassociation line shapes, energy shifts, and rates. \\

The paper is organized as follows: The description of the present model is reported in Section II-IV. The link with previous models, such as the Bohn and Julienne model is discussed in Section V,  where  the various assumptions and approximations are underlined. Application to the cases of Sodium photoassociation in the experimental conditions of Ref. \cite{mckenzie2002} is presented in Section VI. Section VII will summaries and discuss the outlook of the method.

\section{The present  model - preliminary remarks}
The physical process of photoassociation (PA) can  be divided into three steps. {\it Initially} (step I),  a thermal ensemble of cold atoms is in the ground electronic state; the probability to find a colliding pair with  relative motion at energy $E$ and angular momentum $l$ is given by:
\begin{equation}
p_{B}(E)= \frac{(2l+1)\exp(- E/(k_BT))}{Z_{eq}},
\label{boltz}
\end{equation}
where $T$ is the temperature, and $k_B$ is the Boltzmann constant. The partition function $Z_{eq}$ was shown in \cite{koch06} to be well approximated by the classical expression ${Z_{eq}}=Q_{T}V = V(2 \pi \mu k_BT)^{3/2}/h^3$ for a gas of non-interacting pairs in a volume $V$. Further discussion regarding $Z_{eq}$ could be found in \cite{koch06}.\\

The radial wave function which corresponds to the pair is represented with a continuum eigenfunction $\left| {\psi^g_{E,l}(R)  } \right\rangle$ of the single-channel Hamiltonian ${\Op H}_{sc}$ 
\begin{equation}
{\Op H}_{sc} = -\frac{{\hbar ^2 }}{{2\mu }}\frac{{d^2 }}{{dR^2 }} + \frac{{l\left( {l + 1} \right)}}{{R^2 }} + V \left( R \right).
\label{eq:01}
\end{equation}
Here, $\mu $ is the reduced mass of the pair, and $V \left( R \right)$ is the Born-Oppenheimer potential curve for the relevant, here the ground, electronic state.\\

At the second step (step II), the intensity of a CW laser  with frequency $\omega$ is ramped within a time scale that can be varied from hundreds of ns to several $\mu$s. Along  this {\it preparation} period, the continuum states are coupled to one or more bound vibrational levels $v,J$ that belong to one or more excited electronic states $e$. Each of the levels in the excited state, denoted by $\left| {\phi^e_{v,J}(R)} \right\rangle$, may decay spontaneously with a natural free width $ \gamma^{0e}_{v,J}$, which is typically at the order of ns. The field that is applied on this stage varies with time, and so are the states of the total system.\\

Finally, during  the stable  period  of the experiment (step III), the laser reaches its peak constant intensity $I$. For the trap-loss or multiphoton detection scheme which will be described here, the duration of this period is on the order of ms. 

Each of the energy levels of the system at this step can now be written as:
\begin{equation}
\left| {\Psi _j(E)} \right\rangle  = \sum\limits_{e,v,J} {a_{e,v,J}^j \left| {\phi^e_{v,J}} \right\rangle }  + {\sum\limits_{l}{\int {b_j \left( {E} \right)\left| {\psi^g_{E,l} } \right\rangle dE} }} + \sum\limits_{
v',l} {c_{j} \left|{\psi^g_{v',l}}\right\rangle }
\end{equation}
where  we include the possibility of populating bound levels $\left|{\psi^g_{v',l}}\right\rangle $ of the ground state. 
In the present work, we shall consider only this third step, which of course is influenced by the preparation step. Since we shall not treat explicitly the preparation step, and possible transitory effects, a rapid switching on of the photoassociation laser will be assumed, justifying sudden approximation. The density of probability for a transition from initial state $\left| {\psi^g_{E,l}} \right\rangle$ to a final state  $\left| {\Psi_j(E) } \right\rangle$ is given by:
\begin{equation}
P_{(E,l) \to j} = \left| {\left\langle {\Psi _j(E) }
 \mathrel{\left | {\vphantom {{\Psi _j(E) } {\psi^g_{E,l}  }}}
 \right. \kern-\nulldelimiterspace}
 {\psi^g_{E,l}  } \right\rangle } \right|^2 
\end{equation}\\

At section IV it will be shown that a knowledge of $P_{(E,l) \to j}$ is sufficient to extract the experimental parameters of the PA process. An efficient method to compute the initial and the final states, $\left| \psi^g_{E,l}  \right\rangle$ and $\left| {\Psi _j(E)}  \right\rangle$, is the subject of the next section.

\section{Numerical description of the field-molecule system}
The initial state is a thermal ensemble composed of atom pairs, with several  angular momenta $l$. Each pair is coupled to several $J$ partial waves components on the excited state. For simplicity, the equations will be written hereby for one value $l$ and$J$ of the angular momentum. To obtain $P_{(E) \to j}$ one has to compute the initial and final eigenstates of the Hamiltonian without and with the light interaction. In the global grid method used here, this is achieved by representation of the Hamiltonian in a certain basis, and then a diagonalization. The diagonalization step is the most expensive computationally, and it scales as $N^3$ where $N$ is the number of basis function. A selection of an appropriate basis prior to the diagonalization is hence crucial for the efficiency of the method.\\

\subsection{The initial states: $R$-representation of the unperturbed molecule in a box}
Using numerical grid methods, a numerically complete set of of eigenvalues and eigenvectors of the time-independent Schr\"odinger Hamiltonian ${\Op H}^R_{sc}$ is obtained for the isolated molecule, for the ground and excited potentials. This is achieved by using the mapped Fourier grid method (see for example the references \cite{kgrid96,fatal,slava99,nest}). Within the procedure, a unitary transformation ${\Op U}_{mapp}$ is performed on the Hamiltonian eq. (\ref{eq:01}), which is written in equally space grid $R$. The resulted Hamiltonian ${\Op H}^x_{sc} = {\Op U}^\dag_{mapp}{\Op H}^R_{sc}{\Op U}_{mapp}$, is represented on a grid with non-uniform distribution of points in the position space $x$, to take advantage of the phase-space occupancy of the problem. The potential energy operator is written simply as ${\Op V}(x)$, and the kinetic energy operator reads: 
\begin{equation}
{\Op T} =  - \frac{\pi ^2 }{2\mu L_x^2}{\Op J}^{ - 1/2}{\Op D} {\Op J}^{ - 1} {\Op D}{\Op J}^{ - 1/2} 
\end{equation}
$L_x$ is the length of the grid, ${\Op D}$ is the first derivative operator matrix, given by (assuming even number of grid points):
\begin{equation}
{\Op D}_{i,j} = \left( \frac{d}{dx} \right)_{i,j} = \left\{ {\begin{array}{*{20}c}
   {0,i = j}  \\
   {\frac{{\left( { - 1} \right)^{i - j} }}{{\sin \left[ {\left( {i - j} \right)\pi /N} \right]}},i \ne j}  \\
\end{array}} \right.
\end{equation}
and ${\Op J}\left( {x} \right) ={\frac{{dx}}{{dR}}}$ is the (diagonal) Jacobian operator for the transformation. 

A sufficient representation for ${\Op H}^R_{sc}$ by an equally spaced grid demands typically  $N \propto  10^4$ basis functions for the example considered below in section \ref{sec:sodium}. 
As was shown in \cite{slava99}, the use of a mapped grid leads to a reduction of about an order of magnitude in the basis size, to several thousands.\\

A diagonalization of ${\Op H}^x_{sc}$ gives the initial, interaction-free, wave functions and their energies for both electronic states $\left| {\psi^g_{E}(R)  } \right\rangle$ and $\left| {\phi^e_{v,J}(R)} \right\rangle$.
The eigenvalues with $E<0$, correspond to bound, discrete states and the ones with $E>0$ refer to box-normalized quasi-continuum states. The continuum under this picture is therefore tied to the box size from which is was deduced.\\

In order to take account of the  spontaneous emission in the excited channel $e$, an imaginary potential is introduced: \cite{cohen73}
\begin{equation}
{\Op V}_d(R) = i \left( {\frac{2}{{3c}}} \right)^3 \frac{{\left| {\mu_{eg} \left( R \right)} \right|^2 {\omega\left( R \right)} ^3 }}{\hbar }
\end{equation}
where ${ \mu_{eg} \left( R \right)}$ is the transition dipole moment, and ${\omega\left( R \right)} 2\pi \left( {V_e \left( R \right) - V_g \left( R \right)} \right)/\hbar $.

A diagonalization of ${\Op H}^x_{sc}+ {\Op V}_d $ results a complex eigen-energies. The imaginary part of the eigenvalues, $\gamma_{v,J}^{0e}  = {\mathop{\rm Im}\nolimits} \left( {E_{v,J}^e } \right)$ is  interpreted as the lifetime of the level $\left| {\phi _{v,J}^e \left( R \right)} \right\rangle $. 

\subsection{The final states}
\subsubsection{The field-dressed molecule}
A CW laser with constant intensity $I$, detuned by $\Delta_{a}$ relative to the atomic resonance line, introduces a radiative coupling between the ground state and the excited state. After making the rotating wave approximation, the full dressed Hamiltonian under the interaction can be written as:
\begin{equation}
{ {\Op H}_R} = \left[ {\begin{array}{*{20}c}
   {{\Op H}^g_{sc}} & {{\Op \Omega}(R)}  \\
   {{\Op \Omega}(R)} & {{\Op H}^e_{sc}} + \Delta_{a} \\
\end{array}} \right]
\end{equation}
The coupling between the ground and the excited state is introduced by the Rabi coupling operator through the transition dipole moment: ${\Op \Omega}(R) = -A{\Op \mu}_{eg} (R)\sqrt {\frac{I}{{2c\varepsilon _0 }}}$, where ${\Op \mu}_{eg}$ is the spatially dependent transition dipole moment between the two electronic states, and $A$ is an angular factor depending upon the polarization of the laser and the orientation of the molecular axis. In the $R$ representation $\Omega(R)$ is a local, i.e. diagonal, operator. A diagonalization of ${ {\Op H}_R}$ gives the final states $\left| {\Psi _j(E) } \right\rangle$.\\

\subsubsection{Energy eigenstates representation}
For the final states, one has to diagonalize the Hamiltonian describing the interaction of the molecule with the radiation field routinely for many experimental parameters. A direct diagonalization of ${{\Op H}_R}$ even for the mapped grid representation, is inefficient, mainly due to mainly due to the redundancy in the representation of the basis for both potentials. An improved representation for the coupled system is hence crucial. This can be obtained by writing ${\Op H}_R$ in the energy eigenstates of the single channels Hamiltonians ${{\Op H}_{sc}}$. The unitary transformation ${\Op U}_{E-rep}$ is  performed on ${{\Op H}_R}$ to give
\begin{equation}
{\Op H}_{E-rep}={\Op U}^{\dag}_{E-rep} {\Op H}_R {\Op U}_{E-rep},
\end{equation}
the Hamiltonian in the energy representation. The eigenvectors that were obtained as the solutions for the single channel problem serve as the transformation matrices between the two representations, so that:
\begin{equation}
{ {\Op U}_{E-rep}} = \left[ {\begin{array}{*{20}c}
   {\Op U}^g_{E-rep} & 0  \\
   0 & {\Op U}^e_{E-rep} \\
\end{array}} \right]
\end{equation}
where ${\Op U}^{g/e}_{E-rep}$ are the matrices which contain the complete set of eigenvectors for the ground and excited states, $\left| {\psi^g_{E}(R)  } \right\rangle$ and $\left| {\phi^e_{v,J}(R)} \right\rangle$, respectively.\\

The coupled channel Hamiltonian in the energy states picture becomes:\\
\begin{itemize}
\item {for the diagonal part in the excited channel subspace
\begin{equation}
\left\langle { \phi_{v}^e \left( R \right)} \right|{\Op H}\left| {\phi_{v}^e \left( R \right)} \right\rangle  =    E^{0e}_{v} + \Delta_{a} -i\gamma^{0e}_{v}, E^{0e}_{v}  < 0 , 
\label{eq:hdiag_e}
\end{equation}
where  $\Delta_a$ is the detuning from the atomic resonance, and $\gamma^{0e}_{v}$  the natural width for the vibrational level $e,v$.}
\item{for the diagonal matrix elements  in the ground channel subspace:
\begin{equation}   
   \left\langle { \psi_{n}^g \left( R \right)} \right|{\Op H}\left| {\psi_{n}^g \left( R \right)} \right\rangle  = E^{0g}_{n}  .
\label{eq:hdiag_g}
\end{equation}
For clarity, explicit $l$ dependency is ommited. We remark here again that $E^{0g}_{n}$ include both the bound and the quasi-continuum parts of the spectrum. $E^{0g}_{n}>0$ is therefore discrete, for a given box normalization.}\\
\item{ for the non-diagonal part of ${\Op H}_{E-rep}$:
\begin{eqnarray}
  \hbar \Omega _{g,n}^{e,v}  = \langle {\phi^{e}_{v} \left( R \right)} |{\Op H}| {\psi _{n}^g \left( R \right)} \rangle
\label{eq:hcoupling}
 \end{eqnarray}
We can employ the R-centroid approximation for ${\Op \mu}_{eg}(R)$ and write:
\begin{equation}
\hbar \Omega _{g,n}^{e,v} =  {\bar \Omega} \left\langle {{\phi^{e,v} }}
 \mathrel{\left | {\vphantom {{\phi^{e,v} } {\psi_{n}^g }}}
 \right. \kern-\nulldelimiterspace}
 {{\psi_{n}^g }} \right\rangle  \equiv  {\bar \Omega} {\cal \tilde F}_{g,n}^{e,v}
\label{eq:hcoupling2} 
\end{equation}
where now ${\cal \tilde F}_{g,n}^{e,v}$ denotes the Frank-Condon factor between vibrational levels  from the two different electronic states.} 
\end{itemize}

The advantage of the energy representation is the possibility to pick the energy levels that participate in the process and thus reduce the required basis set. We shall consider formally the case of an isolated resonance where the ground state continuum $\left| {\psi_{n}^g \left( R \right)} \right\rangle$ is coupled only to single bound level in the excited state $\left| {\phi_{v}^e \left( R \right)} \right\rangle$. The detuning from the atomic resonance $\Delta_{a}$ could be replaced by $\Delta$, the detuning from the excited bound level. A  generalization to several bound excited states, or even a whole continuum, is straightforward. From this point on, we reduce all the indices for the excited state.\\

By moving to the energy representation we reduce the representation for the coupled channels problem by about one half, to the order of $N \approx 1000$ in the example discussed below in section \ref{sec:sodium}.

\subsubsection{Getting out of the box - from $E$ to $E'$ representation.}
Note that so far, the representation of the continuum is done using a box unity-normalized wave functions. This box-dependent basis for the continuum is arbitrary, and therefore it could be reduced.\\

In order to do so, one had to defined a box-independent Frank-Condon factor.  Box-independent quantities for the continuum are obtained by moving from the box-normalized to the energy-normalized wave functions. The relation between the two is given by:
\begin{equation}
\left| {\tilde\psi_{n}^g \left( R \right)} \right\rangle  = \left( {\frac{{\partial E}}{{\partial n}}} \right)^{ - 1/2} \left| {{\psi}_{n}^g \left( R \right)} \right\rangle 
\label{eq:en_norm}
\end{equation}
where the inverse of the level density, $\frac{{\partial E}}{{\partial n}}$, is given simply, for a large enough box,  by the energy difference between two adjacent quasi-continuum levels, given under the box-normalization constraint. An energy normalized Frank-Condon profile:
\begin{equation}
{\cal F}^{e}(E) =\sqrt{{\frac{{\partial E}}{{\partial n}}}}{\cal \tilde F}_{g}^{e,n}
\label{eq:FC_profile}
\end{equation}
could now be obtained. 
${\cal F}^{e}(E)$ is a box-independent, numerically continuous function of the scattering energy.  In fact, ${\cal F}^{e}(E)$ is a physical measure for the energy dependence of the interaction and it thus the function that one would like to sample for obtaining the minimal energy representation. The alternative basis has to capture the functional energy dependence of the Frank-Condon profile, using the smallest basis. This could be done in principle with any sampling procedure \cite{numerical_recipies}. The technical details regarding the numerical procedure we employed in this work is described at the appendix.\\

Assuming such a minimal representation basis was found, we will denote it  by $\left| g,n' \right\rangle$ for the ground state continuum, for $n'>n_B$, $n_B$ being the number of bound states. The expression for the quasi-continuum diagonal part of the full Hamiltonian (eq.\ref{eq:hdiag_g}) reads now:
\begin{equation}   
   \left\langle { g,n'} \right|{\Op H}\left| {g,n'} \right\rangle  = E^{0g}_{n'},n'>n_B 
\label{eq:althdiag_g}
\end{equation}
and all the other diagonal parts remained unchanged.
The nondiagonal part in the alternative basis is defined through the energy normalized Frank-Condon profile as:
\begin{equation}
\hbar \Omega _{g,n'}^{e} = {\bar \Omega} {\cal \tilde F}_{g}^{e,n'} = {\bar \Omega} {\sqrt{{\frac{{\partial E}}{{\partial n'}}}}}^{-1}{\cal F}^{e}(E).
\label{eq:alt_nond}
\end{equation}
\\

A diagonalization of the Hamiltonian (eqs.[\ref{eq:althdiag_g},\ref{eq:alt_nond}]) gives the set of eigenstates for the coupled Hamiltonian, $\left| {\Psi _j(E) } \right\rangle$, which include components on both channels, with possible contributions from all the bound or free levels. The energy eigenvalues which are obtained are again complex and $\gamma^{j} = \mathop{\rm Im} (E_j)$ is the width of the level $\left| {\Psi _j(E) } \right\rangle$. The transition probability from any initially populated state $\left| g,n' \right\rangle$ to any final state $\left| {\Psi _j(E) } \right\rangle$, is simply:
\begin{equation}
P_{(g,n') \to j} = \left| {\left\langle {\Psi _j(E) }
 \mathrel{\left | {\vphantom {{\Psi _j(E) } {g,n'  }}}
 \right. \kern-\nulldelimiterspace}
 {g,n'  } \right\rangle } \right|^2 
 \end{equation}\\
It is important to notice here that because $\left| {\Psi _j } \right\rangle$ is an eigenstate of the coupled Hamiltonian, a pair that reaches this state at the beginning of the final step of the experiment remains in the same state without any significant dynamics beside an exponential decay with the lifetime $\gamma^{j}$.\\

Table 1 summarizes our procedure so far with the different representations and their typical necessary basis sizes. We started from the Hamiltonian on the equally spaced grid representation. A minimal representation of the problem with such a grid demands a basis of about $N_{R}=10^4$ function. We then used the phase space occupation of the problem to move to a mapped grid, and reduce the basis set in about an order on magnitude to $N_{x}=10^3$, for each of the electronic states. The transformation to the energy representation reduces the excited state manifold. The reduced $E'$ representation minimizes the basis for the ground state, so that finally we could solve the whole coupled channels problem with a basis at the order of $N_{E'}\sim 200$, or even smaller. At this sizes of basis the method become efficient, even compared to the semianalytic perturbative methods. \\

\begin{table}
\begin{tabular}{|c||c||c|}
	\hline
   Representation    & $N_g$ &   $N_e$  \\
	\hline
R - Equally spaced position grid & $10^4$ & $10^4$ \\
x - Mapped position grid & $10^3$&$10^3$ \\
E - Energy eigenstates - boxed &$10^3$& $<10$ \\
E' - Energy eigenstates - non boxed & $\sim 10^2$ & $<10$ \\
	\hline 
	\end{tabular}
\caption{Summary of the different representations used in this paper. $N_g$ and $N_e$ are the typical basis set sizes for the ground and the excited electronic state, respectively, adapted  to the example of section \ref{sec:sodium}.}
\label{param}
\end{table}

In the next section the deduction of the various PA experimental parameters from $P_{(g,n') \to j}$ (and $\gamma^{j}$) is presented.

\section{Extraction of the energy shifts, trap-loss rates, and lineshapes.}
With the formal definitions of the previous subsection we can now move to define
the various physical quantities that are being measured on the system.
The probability to find an atom pair at the beginning of the stable period with energy $E_j$ (which we will denote as zero time) is given by:
\begin{equation}
p(E_j,0) = N_{pair}(0)\sum\limits_{n'} { P_{(g,n') \to j}p_{B}(E_{n'})}=\frac{N^2_{atom}(0)}{2}\sum\limits_{n'} { P_{(g,n') \to j}p_{B}(E_{n'})}
\end{equation}
where $N_{atom}(0)$ and $N_{pair}(0)$ are the number of atoms and atom pairs at time $0$, and $p_B$ is defined above (see eq. (\ref{boltz})). 
As was explained at the last paragraph the only dynamics at the stable period are the exponential decay of the $j$-th level. At later times thus:
\begin{equation}
p(E_j,t) = \exp(-\gamma_j t/\hbar) p(E_j,0)
\end{equation}
so that the total probability density for the whole ensemble is:
\begin{equation}
P_{tot}(t)= \sum\limits_{j} p(E_j,t) = \frac{N^2_{atom}(0)}{2} \sum\limits_{n',j} \exp(-\gamma_j t/\hbar) { P_{(g,n') \to j}p_{B}(E_{n'})} 
\end{equation}
The observed reaction rate $v(t)$ at any given time  is the change of $P_{tot}(t)$ versus time:
\begin{equation}
v(t) = -\frac{dP_{tot}(t)}{dt} = \frac{N^2_{atom}(0)}{2\hbar} \sum\limits_{n',j}\gamma_j \exp(-\gamma_j t/\hbar) { P_{(g,n') \to j}p_{B}(E_{n'})}  \equiv \frac{N^2_{atom}(0)}{2\hbar} \left\langle {\gamma(t)} \right\rangle 
\end{equation}
The rate could be thus  interpreted as the thermally averaged lifetime. 
On the other hand, the experimental definition of the diatomic reaction rate for a sample containing $N_{atoms}$ is:
\begin{equation}
v(t)= - K \frac{N_{atoms}^2(t)}{V}.
\end{equation}
Now, a comparison between the two expressions yields the relation between the calculated and experimental rate constants, which holds for any given time. Specifically for t=0 one gets:
\begin{equation}
K(I,\Delta) = \frac{V \left\langle {\gamma(0)} \right\rangle}{2Z_{eq}} = \frac{\left\langle {\gamma(0)} \right\rangle}{2Q_T}
\label{our_K}
\end{equation}
In a fashion similar to the experimental, a calculation of $K$ as a function of
the detuning yields the lineshape. The peak value of $K$ as a function of 
$\Delta$, and the value of the detuning at the peak are termed as the maximum reaction rate $K_{max}$ and the energy shift $E_1$ for a given intensity and temperature, respectively.
\\

Before moving forward, we remark that the model is non perturbative with respect to the
laser intensity. It is also well oriented to dynamical, time dependent-type 
calculations. Furthermore, a generalization of the method to multiple coupled 
continuum manifolds, several bound states, or to include hyperfine structure,
is strightforward. Its drawback is mainly being a little less efficient
computationally, mainly due to the expansive diagonalization step. Nevertheless, this could be overcomed, 
especially for calculating lineshapes and saturation effects, as we will show at the 
 application of section VI. But, before doing so, we will present briefly in the next subsection the necessary expressions of the method of Bohn and Julienne\cite{bohn99} (B $\&$ J) to which our expression (\ref{our_K}) will be compared to.

\section{Adjustment of Bohn and Julienne model to the grid methods framework}
In this subsection we present the method of B $\&$ J \cite{bohn99}, and adjust it
to take advantage of the available grid methods.
We will limit the treatment to the simple case of a single isolated resonance embedded in a continuum.
The scattering-theory-based method defines a scattering amplitude $S_{E}$ for a
pair of atoms in the center of mass framework. The square of $S_{E}$ is referred to as the
probability to initiate in a unbound state with kinetic energy $E$ and be lost
at infinite time, due to a light-induced coupling to a bound molecular state.

According to Fano\cite{bohn99,fano61}, $|S_{E}|^2$ is given by:
\begin{equation}
|S_E|^2 = \frac{{\gamma \Omega_E }}{{\left( {E - \Delta - E_1 } \right) - \left(
{\frac{{\gamma  + \Omega _E }}{2}} \right)2 }}.
\end{equation}
here $\Delta$ is the detuning of the light frequency of the resonance bound state's
energy , and $\gamma$ is the natural linewidth of the bound level. $E_1$ and
$\Omega_E$ are the energy shift and the stimulated absorption/emission rate,
respectively. With the help of second order perturbation theory it was shown
in \cite{fano61} that:
\begin{equation}
E_1  = {\cal P}.\int {dE'\frac{{\left| {\Omega _{E'} } \right|^2 }}{{\Delta  -
E'}}}
\label{fanoshift}
\end{equation}
where ${\cal P}.$ stands for `the principle part of', and $\Omega_E$ is the
dipole moment coupling that was defined above (see eq.(\ref{eq:hcoupling})).

The rate of loss from the system is given by performing a velocity thermal
averaging of $|S_E|^2$, for all the initial scattering energies, and with respect to all
the relevant partial waves, to obtain the observed rate loss:
\begin{equation}
K = \sum\limits_l {\frac{{\pi v}}{k}\int {p_B \left( {E^l } \right)\left| {S_E^l } \right|^2 dE} } 
\label{KBohn}
\end{equation}
where the integral over the energy indicates for a thermal averaging over the Boltzmannian distribution $p_B$, defined above (see eq.\ref{boltz}). The superscript for $S$ denotes
the rotational quantum number for the partial wave $l$, and $v$ and $k=\sqrt(2\mu E)/\hbar$ are the relative velocity and the wavenumber of the atom-pair. 

It is important to comment here, that in order to calculate the energy dependent $\Omega_E$ 
for eqs. (26-28) the energy dependency of the continuum has to be calculated [see eq. (\ref{boltz})]. The common use of local solvers like the Numerov propagator \cite{numerical_recipies}
for that purpose made Fano's formalism impractical computationally.
An energy independent $\Omega_E$ was assumed 
for demonstrative purposes in Ref. \cite{fano61}. Obviously, as one can see from figure 3
this is far from being adequate to describe molecular systems in any other
circumstances.

Using the ideas of Du et al. \cite{du89}
Bohn and Juliene combined the couple of regular and irregular solutions of eq.(\ref{eq:01}) to write the energy shift $E_1$ as:
\begin{equation}
E_1  = -2\pi \int\limits_0^\infty  {{^r\Omega _E} \left( R \right)}dR
\int\limits_0^R {{^i\Omega _E} \left( {R'} \right)}dR'
\end{equation}
where the superscripts denote the regular and irregular part of the continuum
wavefunction for $E\to 0$, and $^{r/i}\Omega_E(R)$ is understood to be the
integrand of the integral of eq.(\ref{eq:hcoupling}).
As noted in \cite{du89,portier2006}, this alternative expression could be viewed as a
way to write the Green kernel ${\Op G} = \left({\Op H}-E\right)^{-1}$ in the
position representation instead of the energy space used in Fano's
formalism.\\

To reduce even further the effort in the computation of $\Omega _E$,
the assumption of a slowly varying continuum with respect to $E$ is employed, i.e., $\lim_{E\to 0}\frac{d\Omega _E}{dE} \ll p_B(E\to 0)$. Under these conditions, the rate is assumed to be influenced by the on-resonance light interaction.

The continuum is represented now by only a single continuum state and the thermal averaging in
eq.(\ref{KBohn}) can be avoided. As shown in\cite{Bohn99}, this leads to:
\begin{equation}
K = \frac{h}{2\mu k}\frac{4\gamma\Omega(E_R)}{(\gamma + \Omega(E_R))2}
\label{Kapp}
\end{equation}
where now the coupling $\Omega_E$ (defined above in eq.(\ref{eq:hcoupling}))  is taken only on the resonance energy $E_R$.

An analytic alternative way for calculating the energy dependence of $\Omega$
was suggested in Ref. \cite{crubellier2006}. In the latter work, strong energy dependence,  was found  in some cases, for instance $^{85}$Rb.

In this paper, we adopt the advantages of the grid methods in the
E-representation, and calculate the explicit dependency of $\Omega(E)$ directly from the continuous energy normalized
Frank-Condon profile with any desired accuracy for any energy range. This allows us to calculate $E_1$ directly using Fano's formalism, without a need to have the irregular solution of the single channel Schrodinger equation. As noted in \cite{portier2006} the two expressions give the same results.\\

At the next section a comparison between the various expressions will be demonstrated by a concrete application.

\section{Application}
\label{sec:sodium}
\begin{figure}[htbp]
    \includegraphics[width=0.95\linewidth]{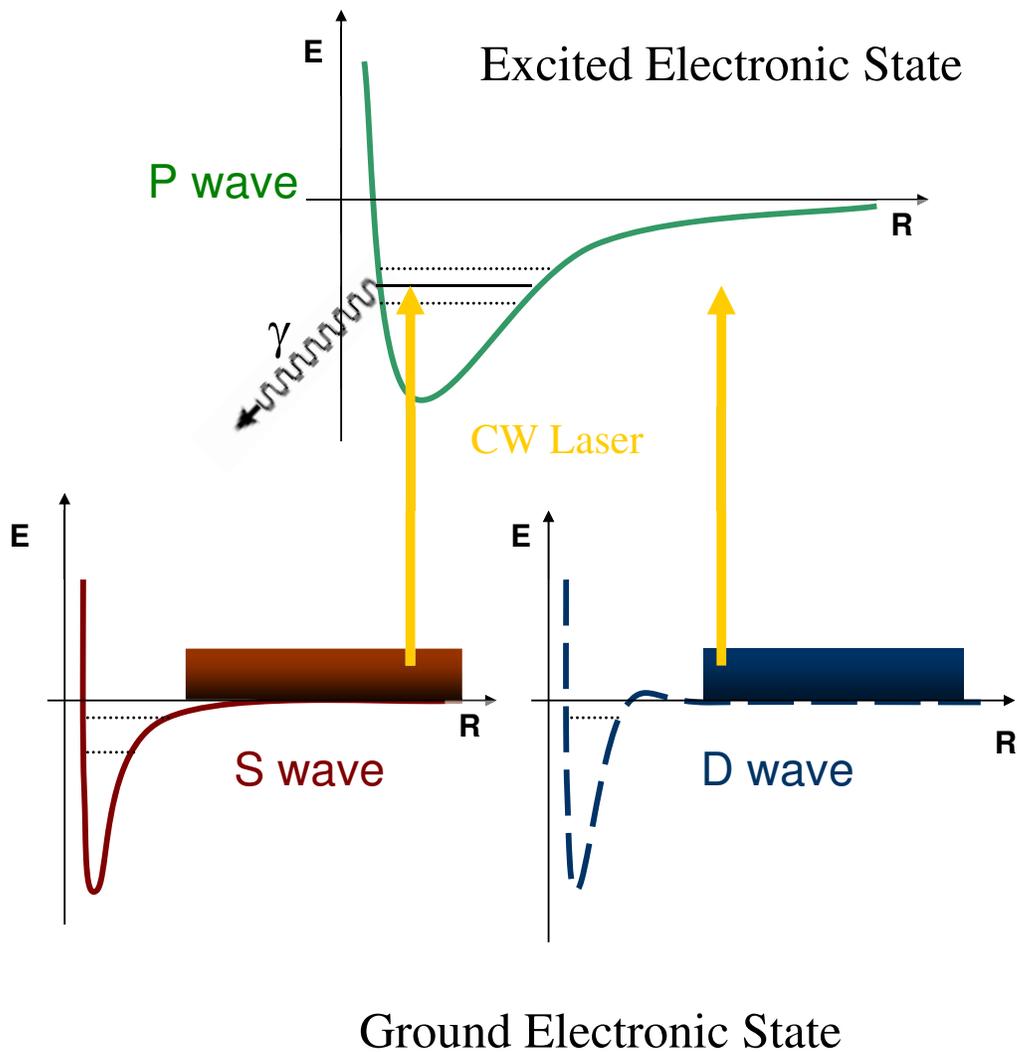}\hfill%
    \caption{A schematic view of the CW laser PA experiment. Atoms pairs colliding in $S\, (l=0)$ and $D\, (l=2)$ partial waves on the ground electronic state, are coupled by a laser with constant frequency and intensity to the $v=135, P\, (J=1)$ bound level in the excited electronic state. The nearby bound levels on the excited state, as well as few  bound levels within the ground electronic state could be coupled as well. The natural lifetime of the bound state $\gamma$ is $2\pi\times$ 18.36 MHz.}
    \label{fig:scheme}
\end{figure}

The example chosen for numerical application will be the on-resonance photoassociation  of the $v$=135 , $J$=1 level of
Na$_{2}$ A$^{1}\Sigma_{u}^{+}$, in the experimental conditions of the NIST experiment \cite{mckenzie2002}. The initial state is a pair of colliding atoms with $l=0$ or $2$ (see figure \ref{fig:scheme}). 
The calculations have been performed  using the same potentials as in  Ref.\cite{naidon2006}. The position of the repulsive wall in the ground state X$^{1}\Sigma_{g}^{+}$ has been slightly modified, so that the binding energy of the last levels is $E(v=65)=0.013cm^{-1}$ for $l$=0 and $E(v=64)$=0.25cm$^{-1}$ for $l$=2. The experimental values for these levels are $E(v=65)=0.013$cm$^{-1}$ and $E(v=64)=0.25$cm$^{-1}$, respectively. In the excited curve, the level $v$=135 is bound by 49.23 cm$^{-1}$, the two neighboring levels being $v$=134 at -52.95 cm$^{-1}$ and  $v$=136 at -45.76 cm$^{-1}$. Note that in experiment the binding energy of the resonant level is 43 cm$^{-1}$. The natural width is set to $\gamma$ = 2 $\pi \times$ 18.36 MHz to match the experiment. \\

\begin{figure}[htbp]
\epsfig{file=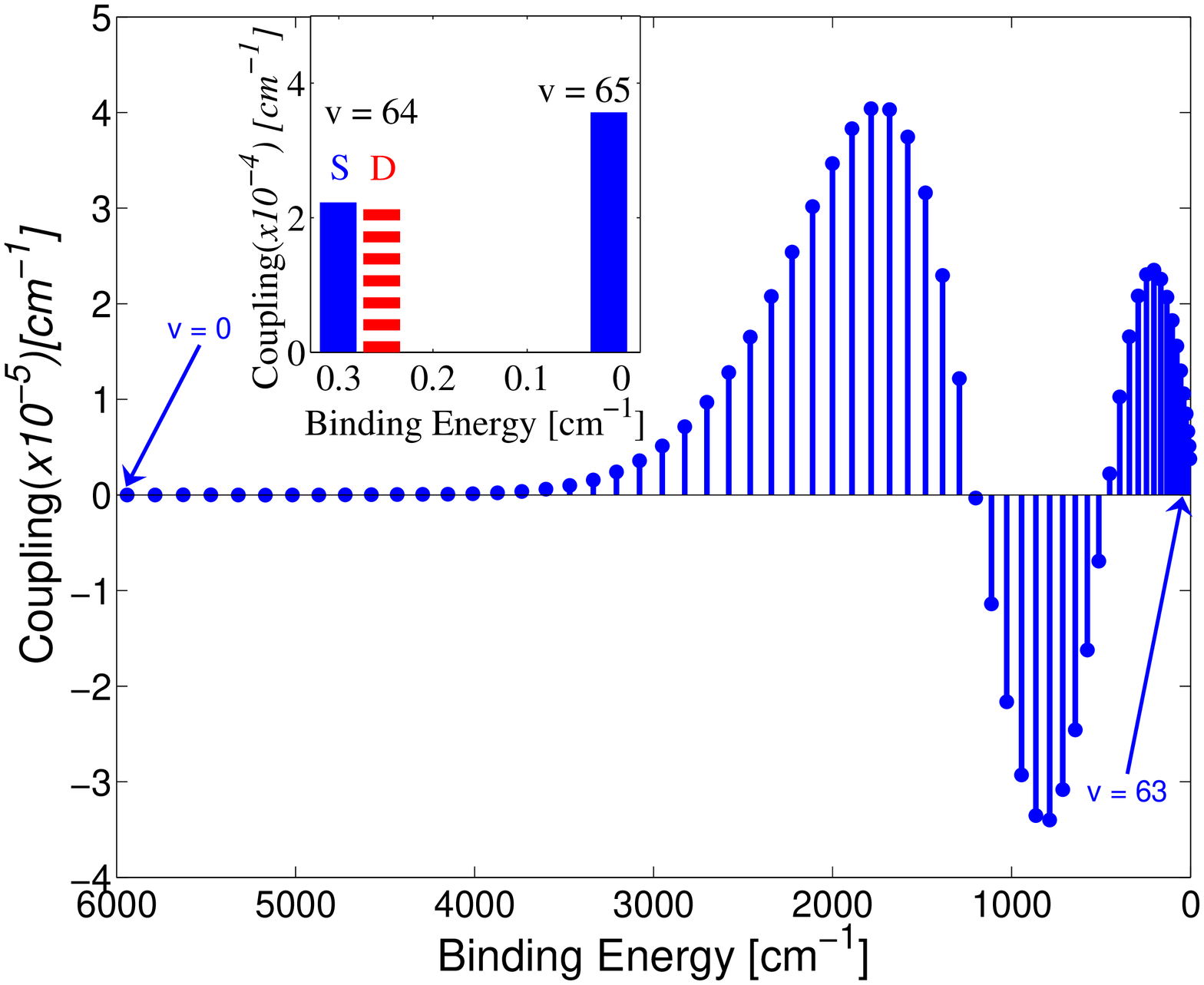, angle=-90, width=\linewidth}
    \caption{Variation of the coupling term between the level $v$= 135 of Na$_2$ A$^1\Sigma_u^+$ and the vibrational levels of the Na$_2$ ground state vs. the binding energy. The intensity of the CW laser is $I$=1 W/cm$^{-2}$. The rotational quantum number is assumed to be either 0 or 2. The first 64 levels are shown. Inset: the same for the upper levels for a rotational quantum number $l$= 0 (full rectangles, $n$=64 and $n^0_B$=65) and $l$=2 (striped rectangle, $n=n^0_B$=64).}
    \label{fig:coupling_bound}
\end{figure}

Figure \ref{fig:coupling_bound} presents the coupling profile, i.e., the Frank Condon factors multiplied by the Rabi frequency (see eq. \ref{eq:hcoupling2}), for various bound levels $n$ of the ground state coupled to one  level (v=135)  of the excited state  at intensity of 1 W/cm$^2$. In the main frame $n$ varies from 0 to 64 and oscillations in the coupling term are visible as the Condon point is moving to large internuclear distances .  The  FC factors are identical for both partial waves when $n \le 64$ . The $n = 65$ level is bound only for the S wave.\\

At the upper and the lower panels of figure \ref{fig:coupling_cont} the coupling profile for the continuum part of the spectrum is shown, for both $S$ and $D$ partial wave, respectively. The left and the right panels show the same profile in logarithmic and linear scale. The last bound level $n = 65$ for the $S$ wave, is shifted for the $D$ wave and  appears as a shape resonance with a well defined width in the $D$ continuum. For the continuum part of the spectrum, the Condon point is energy independent. The structure of the coupling profile of the continuum is a signature of the energy dependence of the continuum wave functions\cite{koch06}. The rotational barrier on the $D$ potential surface makes the $D$ profile to vanish more strongly near the threshold than does the $S$ profile.\\

In the next subsections the results of grid method will be presented in three steps: First, the light induced energy shift will be calculated and discussed. Then, the results for the rest of the PA observables will be presented in the low intensity regime, where the perturbative method should be adequate. Finally, the results for in the high intensity regime, where the appearance of saturation effect are explored.\\

\subsection{Light Induced Energy Shifts} 
\begin{figure}[htbp]
\epsfig{file=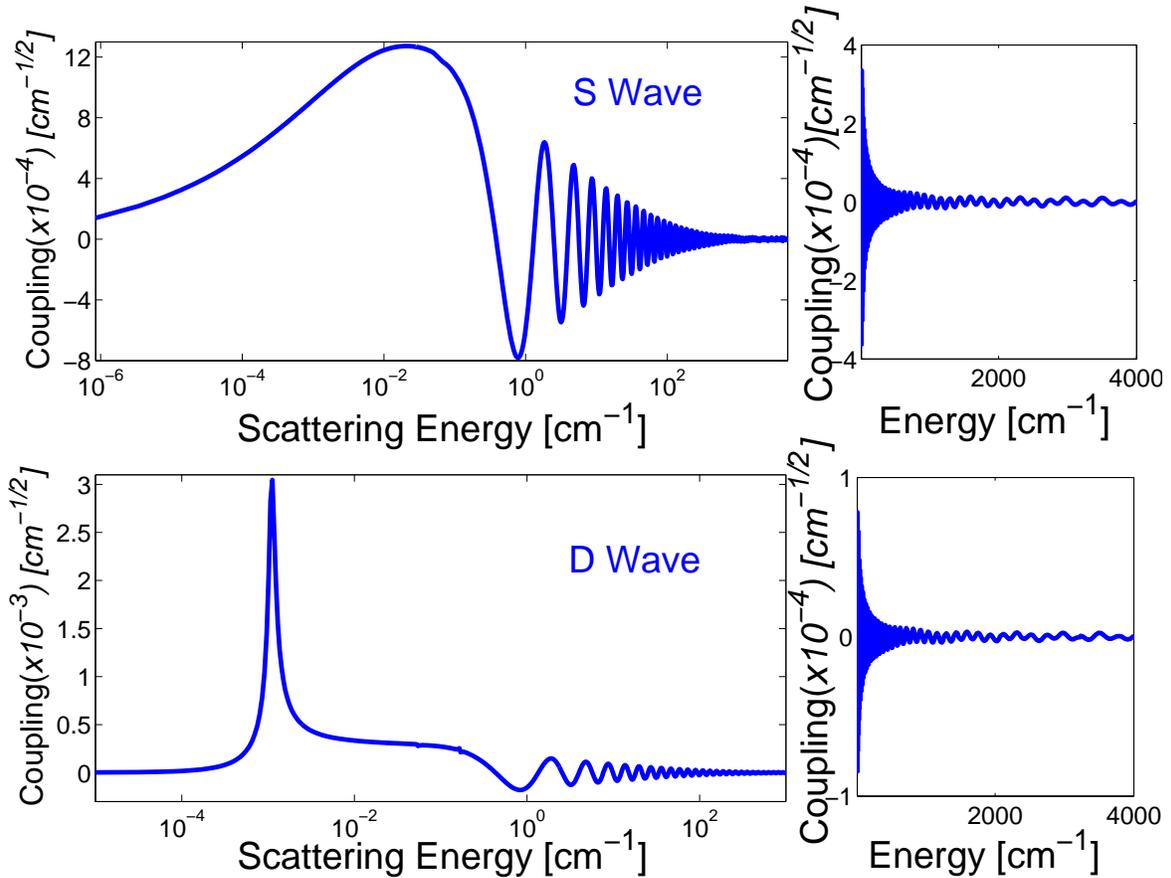, angle=-90, width=\linewidth}
    \caption{Left: Variation of the coupling term between the level $v$= 135 and the energy normalized  $S$ (upper) and $D$ (lower) continuum  levels  of the Na$_2$ ground state vs. the scattering energy at logarithmic scale. The intensity of the CW laser is $I$=1 W/cm$^{-2}$. Right: the same as the left panels, now in a linear scale.}
    \label{fig:coupling_cont}
\end{figure}

Using Fano's perturbative expression for the energy shift (eq.\ref{fanoshift}) and the right panels of figure \ref{fig:coupling_cont} it is understood that the energy scale which is needed for the calculation of the energy shift is larger in many order of magnitude then the thermal energy. (see the appendix for a numerical example). On the other hand, as will be shown below, the lineshape is influenced only by the coupling profile in the vicinity of the thermally populated energy. This separation of scales can help us furthermore in reducing the basis set, depending upon which physical quantity one would like to calculate. \\

\begin{figure}[htbp]
\epsfig{file=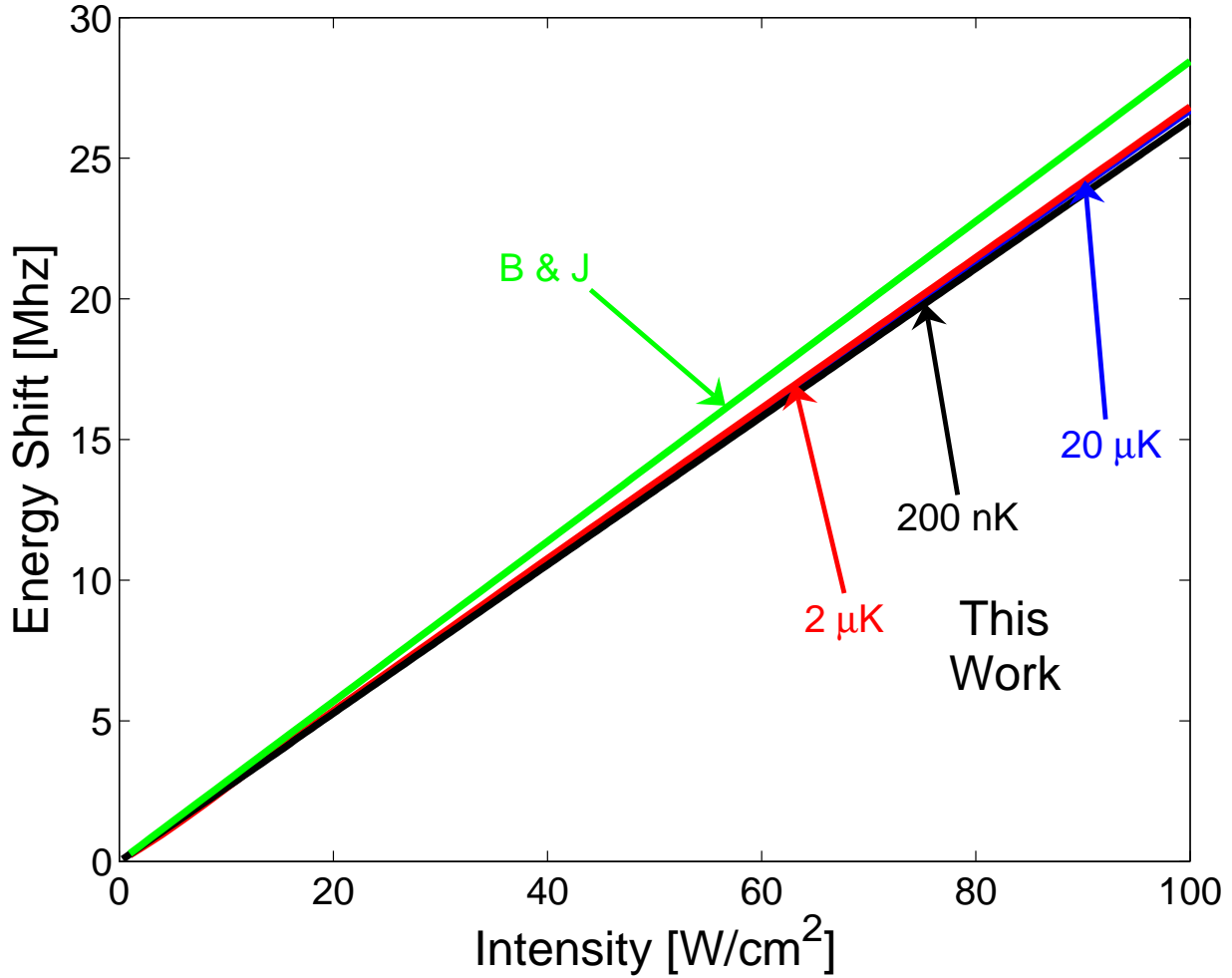, angle=0, width=\linewidth}
    \caption{Energy shift as a function of the intensity. The values for the perturbative method are in green, irrespective on the temperature. In red, blue (barely visible) and black: non perturbative method, for $2\mu K$, $20\mu K$, and $200 nK$, respectively. Only the interaction with the $S$ continuum is considered.}
    \label{fig:eshiftS}
\end{figure}

\begin{figure}[htbp]
\epsfig{file=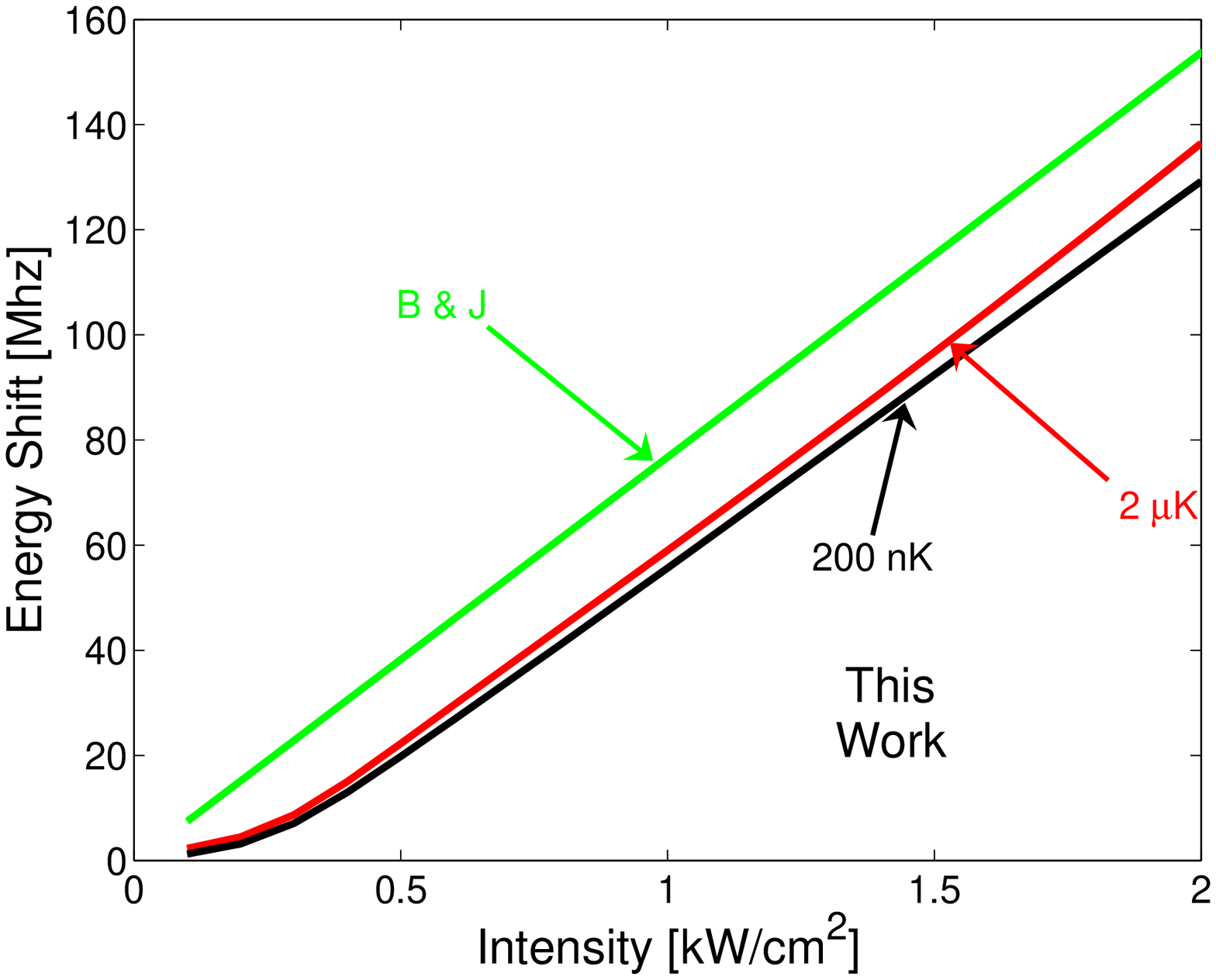, angle=0, width=\linewidth}
    \caption{The same as in figure \ref{fig:eshiftS}, with the full interaction: $S$ and $D$ continuum manifolds and bound states.}
    \label{fig:eshiftall}
\end{figure}

The light induced energy shift for intensities up to 100W/cm$^2$, are presented in figure \ref{fig:eshiftS}. The energy shift was calculated using the $S$ continuum part of the ground state manifold. The green line is the energy shift that was calculated using the modified Bohn and Julienne perturbative method and is temperature independent by definition. In red, blue and black, we present the energy shift using our method for $T = 20 \mu K,2 \mu K$ and $200nK$, respectively. All the curves are linear as expected. A difference exists between the energy shifts, but it is too small to be measurable. The difference in the slopes between the present calculations and the perturbative method is also within the experimental accuracy. An important result we observed in our calculation is that the perturbative linearity of the energy shift with respect to the intensity was conserved for any feasible intensity that was checked (up to $100kW/cm^2$ and higher). Another feature of the perturbative limit is the additivity of the contributions from different partial waves with respect to the bound and the continuum. Figure \ref{fig:eshiftall} presents the energy shift for the full interaction, i.e., bound and continuum part of the spectrum for both $S$ and $D$ partial waves. No deviations from the perturbative treatment were found with respect to the additivity.\\

\subsection{Low Intensity regime: the perturbative limit}  

\begin{figure}[htbp]
\epsfig{file=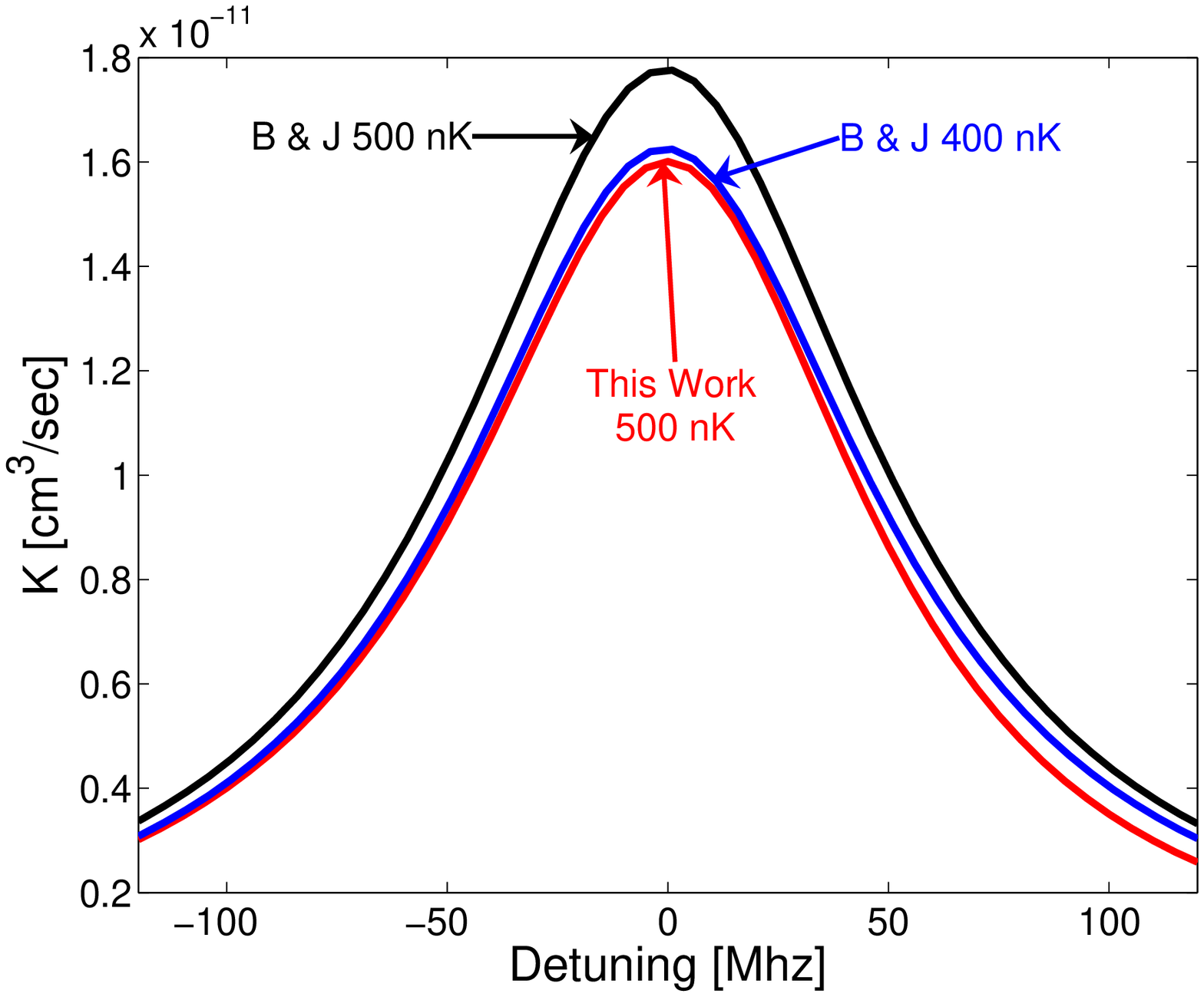, angle=0, width=\linewidth}
    \caption{A typical lineshape, i.e., rate of loss vs. the detuning, for the low intensity regime. (red) The non-perturbative method for 500nK. (black and blue) The perturbative method, for 500nK and 400nK, respectively. All the lines correspond to full energy averaged calculations, and are shifted to be centered at zero detuning, for comparison.}
    \label{fig:line_low}
\end{figure}

\begin{figure}[htbp]
\epsfig{file=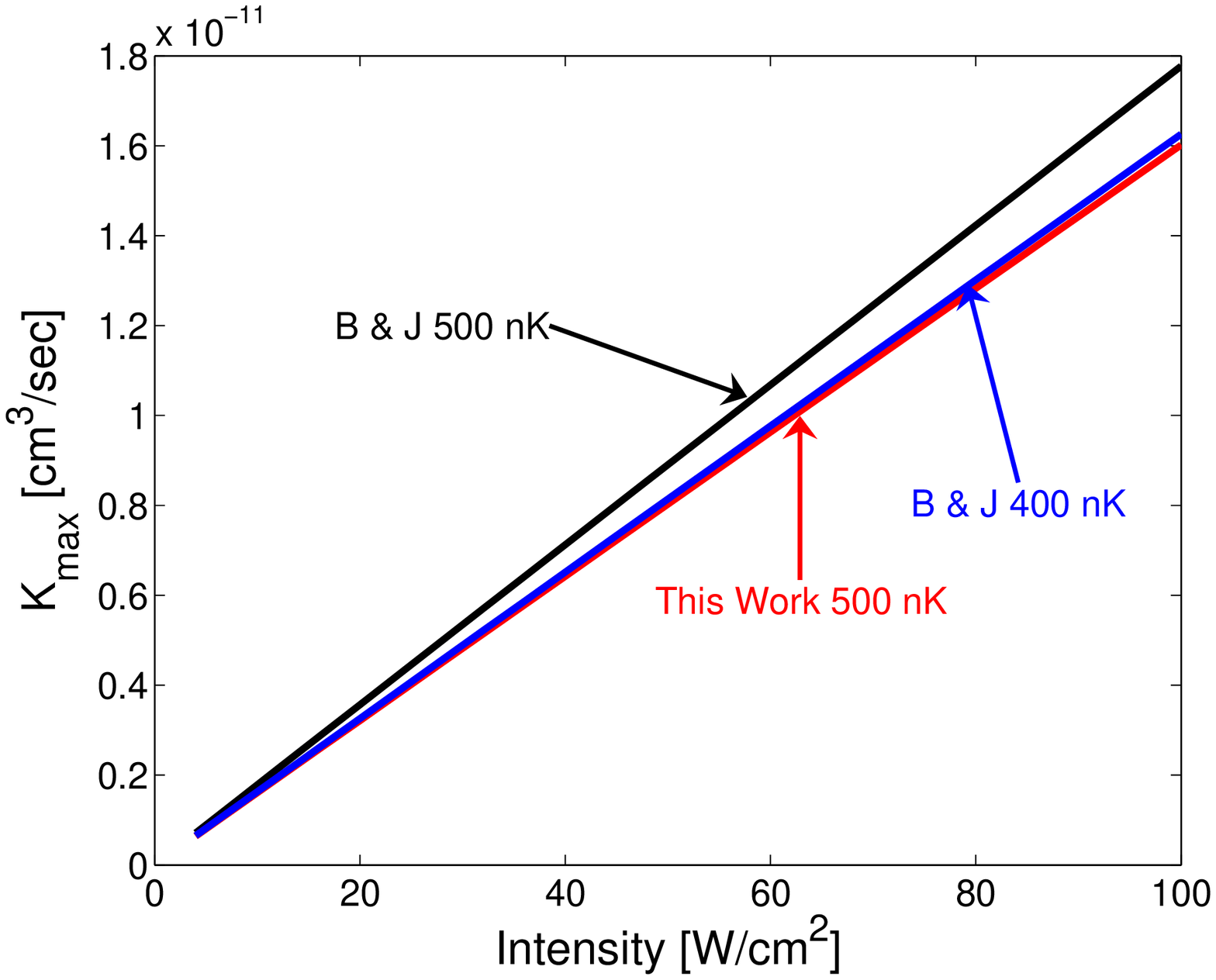, angle=0, width=\linewidth}
    \caption{The intensity dependency of the peak loss rate values for a given lineshape ($K_{max}$) for the non-perturbative and perturbative methods. Colors are the same as in figure 5. All lines correspond to thermally averaged calculation for the assigned temperatures.}
    \label{fig:K_vs_Ilow}
\end{figure}

Saturation effects for the case of photoassociation in a sodium condensate were not observed up to 1.2kW/cm$^2$. 
A typical lineshape for an intensity of 100W/cm$^2$, well below the saturation limit, is presented in Figure \ref{fig:line_low}. The lines were all centered to zero detuning. The red line presents the rate versus the detuning for our method for T=500 nK. In black line the correspondent result for the Bohn and Julienne method with the same temperature is shown. The two lines show the same width but differ at the peak height by less then $10\%$, small comparable to the experimental accuacy. Moreover, as can be seen from the blue line in the figure, a small change of the temperature for the perturbative method (to $400nK$)reproduces exactly the same lineshape as in our method. Figure \ref{fig:K_vs_Ilow} presents with the same colors of figure \ref{fig:line_low} the results for the maximum rate  lineshape $K_{max}$ as a function of the intensity. All the curves are linear, as expected for the perturbative regime. Again, the agreement between the two methods is within the experimental accuracy.\\

\begin{figure}[htbp]
\epsfig{file=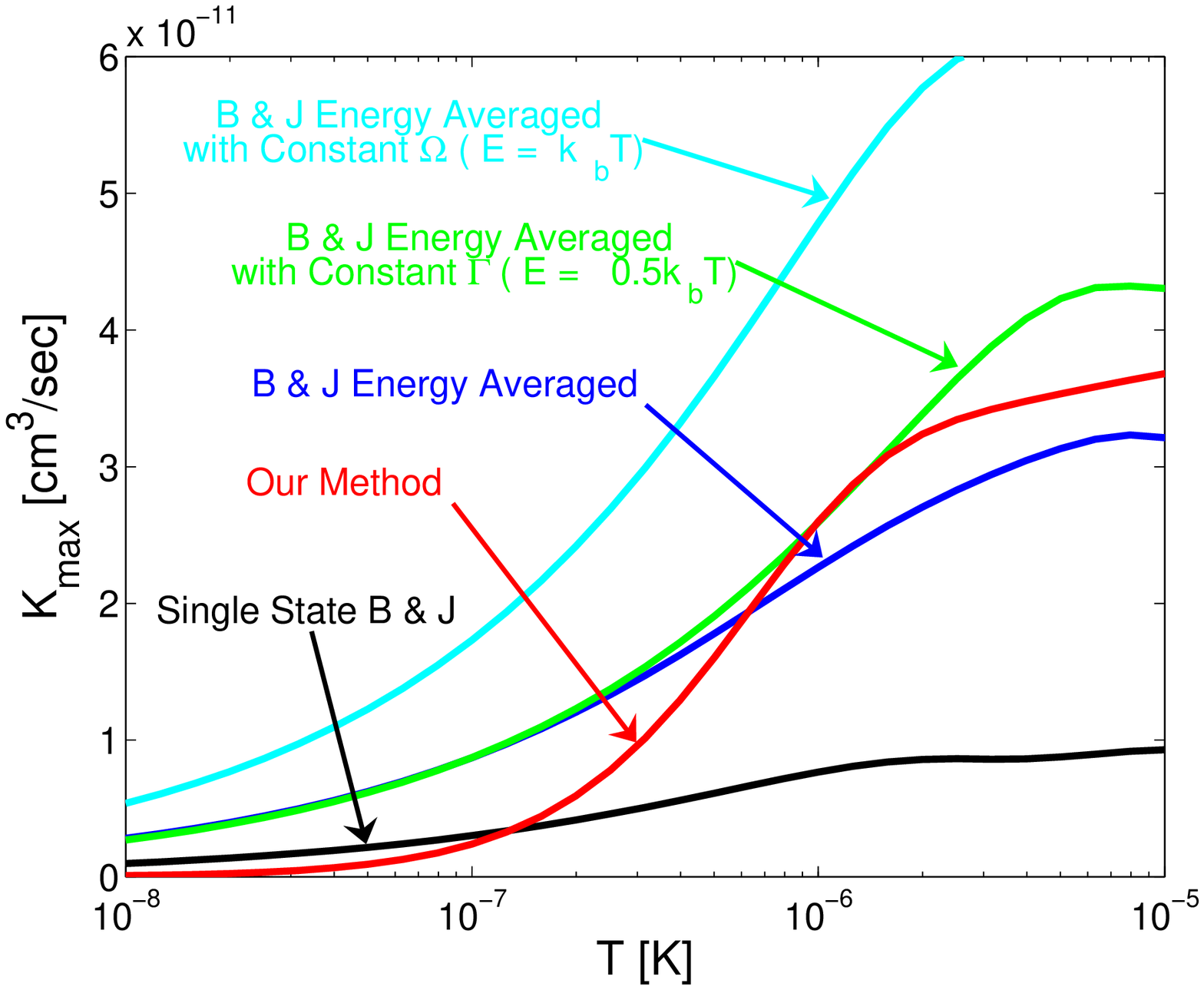, angle=0, width=\linewidth}
    \caption{The peak loss rate for a given lineshape, $K_{max}$, vs. the temperature for the three models. Red: this paper's model, Blue: the full thermally averaged perturbative, and Black: the single continuum energy approximation, see eq. (\ref{Kapp}) in the text. Cyan and Green: the perturbative model with energy averaged but with a constant coupling $\Omega$, taken at $k_bT$ and $0.5k_bT$, respectively.} 
    \label{fig:T_dep}
\end{figure}

We have to remark here that, to our best knowledge, an extensive study of the temperature dependency of photoassociation lineshapes has never been performed. Figure \ref{fig:T_dep} presents the temperature dependency of $K_{max}$ for intensity of 100W/cm$^2$. The red and blue lines present the values for the thermal averaged rates of expressions (\ref{our_K},\ref{KBohn}). The difference between the methods at the most common temperature for PA experiments in condensates, i.e., a few hundreds of nano Kelvins to micro Kelvins, is at the order the experimental uncertainty. The values deviate significantly from each other at relatively high ($T>10\mu K$) and low ($T<100nK$)temperatures. The black line in figure \ref{fig:T_dep} represents the values obtained from the perturbative method without energy averaging (see eq. \ref{Kapp}). A single state approximation for the rate seems to be inadequate, at least for the description of most of the temperature dependence of th
 e trap loss. In addition, we checked the results obtained applying the perturbative model (\ref{KBohn}) with thermal energy averaging but assuming a constant coupling $\Omega(E)$. The cyan and green lines show the values for $\Omega(E)$ taken at $k_bT$ and $0.5k_bT$, respectively. The constant coupling approximation fits perfectly to the fully thermally averaged model for the low temperature regime with $E=0.5k_bT$. Two factors play a role in this situation: the coupling profile (see fig. \ref{fig:coupling_cont}) and the Boltzmann thermal distribution (see eq. \ref{boltz}). The first increases with $E$ while the last decreases. From the results it seems that in our case, the compensation of the two factors takes place at $E$=0.5$K_BT$. Above this temperature, the thermal width starts to play a role, and the constant coupling approximation is no longer valid.\\

\subsection{Non Perturbative regime - saturation effects}   
\begin{figure}[htbp]
\epsfig{file=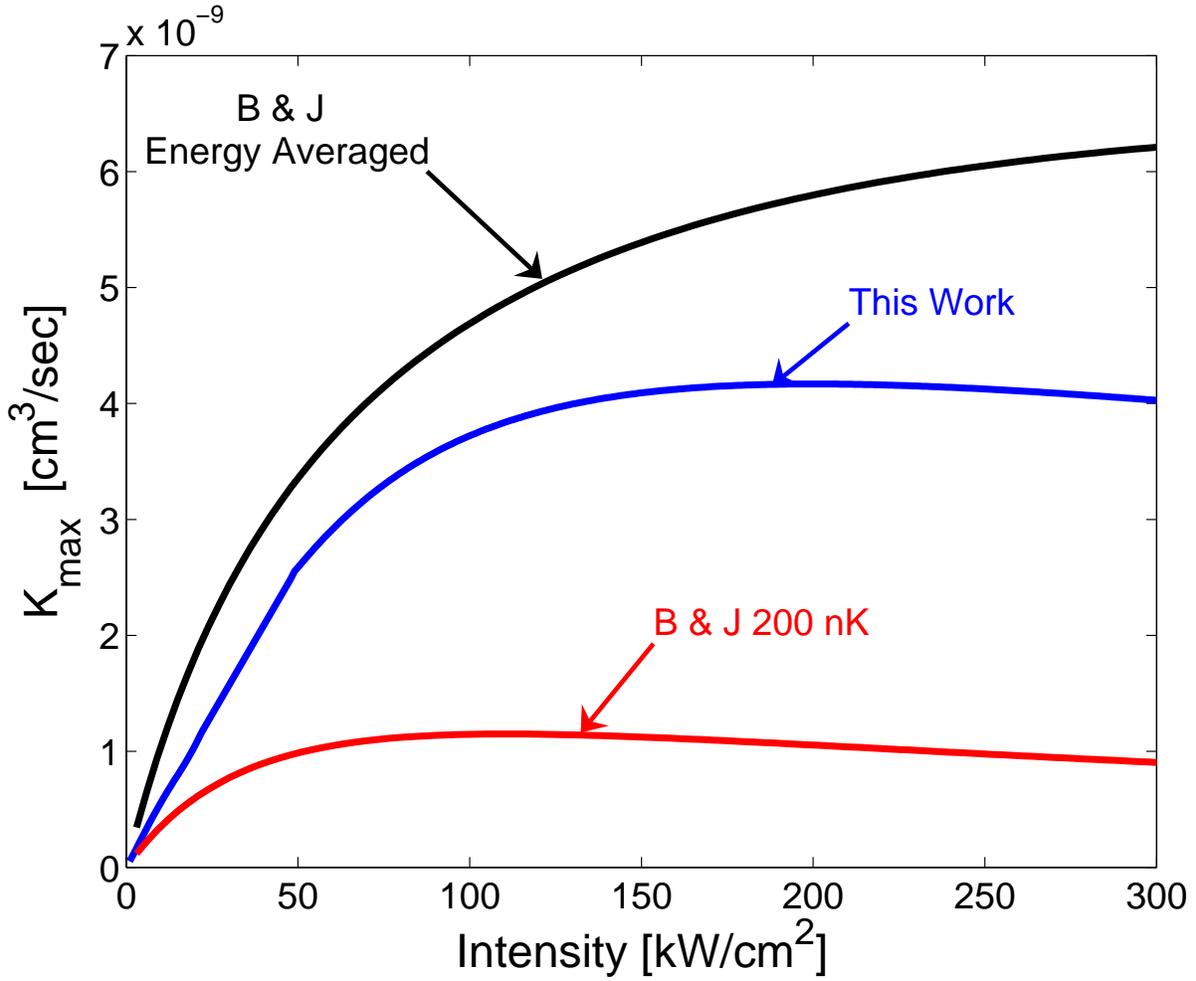, angle=0, width=\linewidth}
    \caption{Same as figure (\ref{fig:K_vs_Ilow}), now for high intensities and $T=200nK$, a typical condensate temperature. Blue and Black and the non perturbative and the perturbative fully thermally averaged models. Red: a single continuum level at $200 nK$.}
    \label{fig:K_vs_Ihigh_nK}
\end{figure}

We now move to explore the regime of high intensity. In figure \ref{fig:K_vs_Ihigh_nK} we present the intensity dependence of $K_{max}$ for $I$ up to 300 kW/cm$^2$ . The blue, black and red lines are the results for our method, the thermally averaged, and the non averaged perturbative method, respectively. The temperature is $200 nK$,  typical for a condensate. All the lines show saturation behavior, but the saturation limit is not the same. The results for the single energy perturbative approximation are significantly lower then the other two methods. The intensities for reaching saturation are way too high to be observed experimentally, and, indeed, were never observed. The validity of the two-body model does not hold for such intensities, and therefore the significance of the results is mainly in predicting a saturation effect for other species where the saturation is reached at lower intensities, e.g., cesium\cite{kraft2005}, lithium\cite{prodan2003} or strontium \cite{nagel2005}.\\

\begin{figure}[htbp]
\epsfig{file=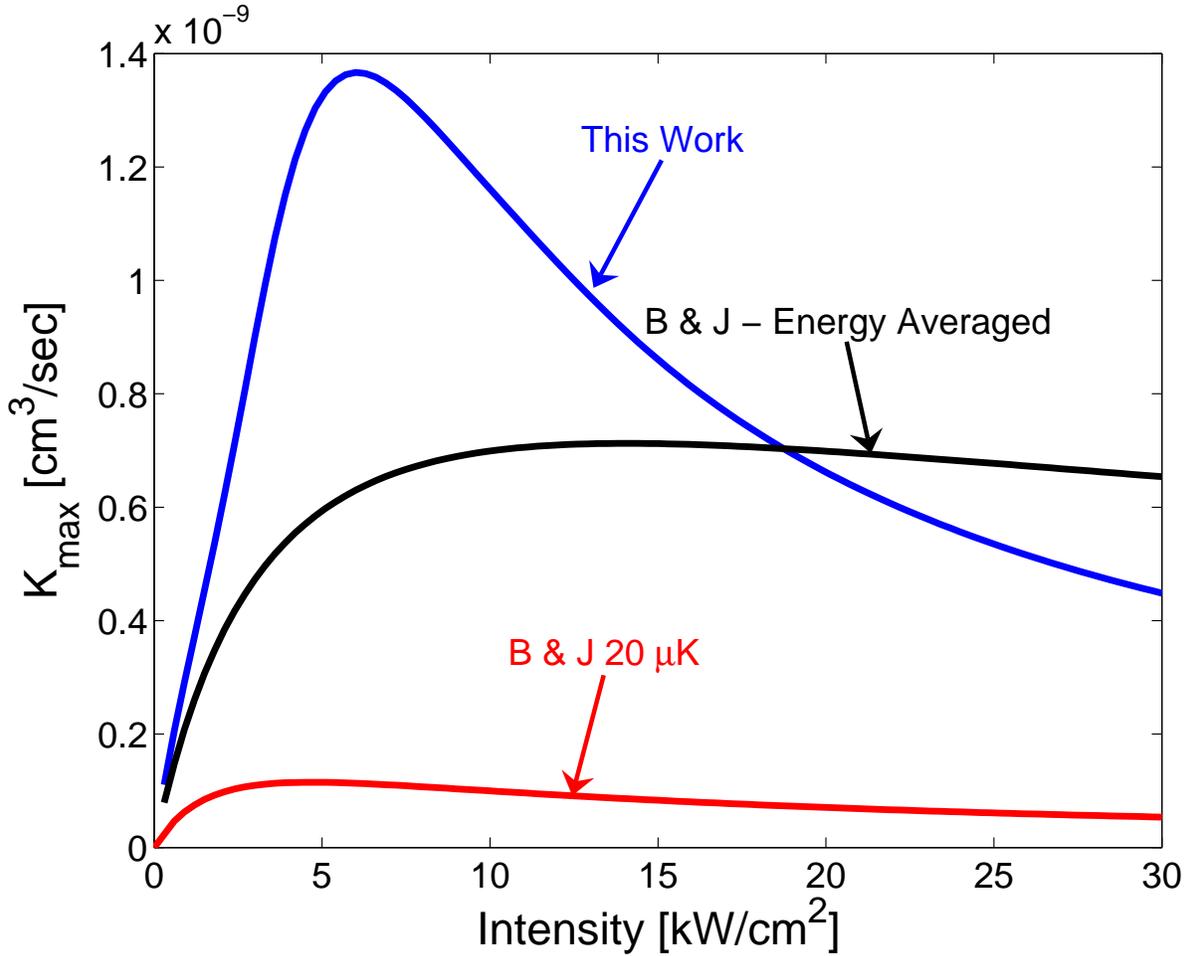, angle=0, width=\linewidth}
\caption{Same as figure \ref{fig:K_vs_Ihigh_nK}, for $T=20 \mu K$, a typical temperature for a MOT.}
    \label{fig:K_vs_Ihigh_muK}
\end{figure}

Figure \ref{fig:K_vs_Ihigh_muK} presents the same calculations as figure \ref{fig:K_vs_Ihigh_nK}, but for $T = 20\mu K$, a typical temperature for a MOT. At a MOT temperature saturation is reached at lower intensities for all the models. The non-perturbative method presents an inverted region in which the peak rate constant decrease with the light intensity. This prediction has never been observed, and it might be feasible experimentally for species other then Sodium. The saturation for the case of cesium in a MOT may give clue to this inverted behavior\cite{kraft2005}\\

\begin{figure}[htbp]
\epsfig{file=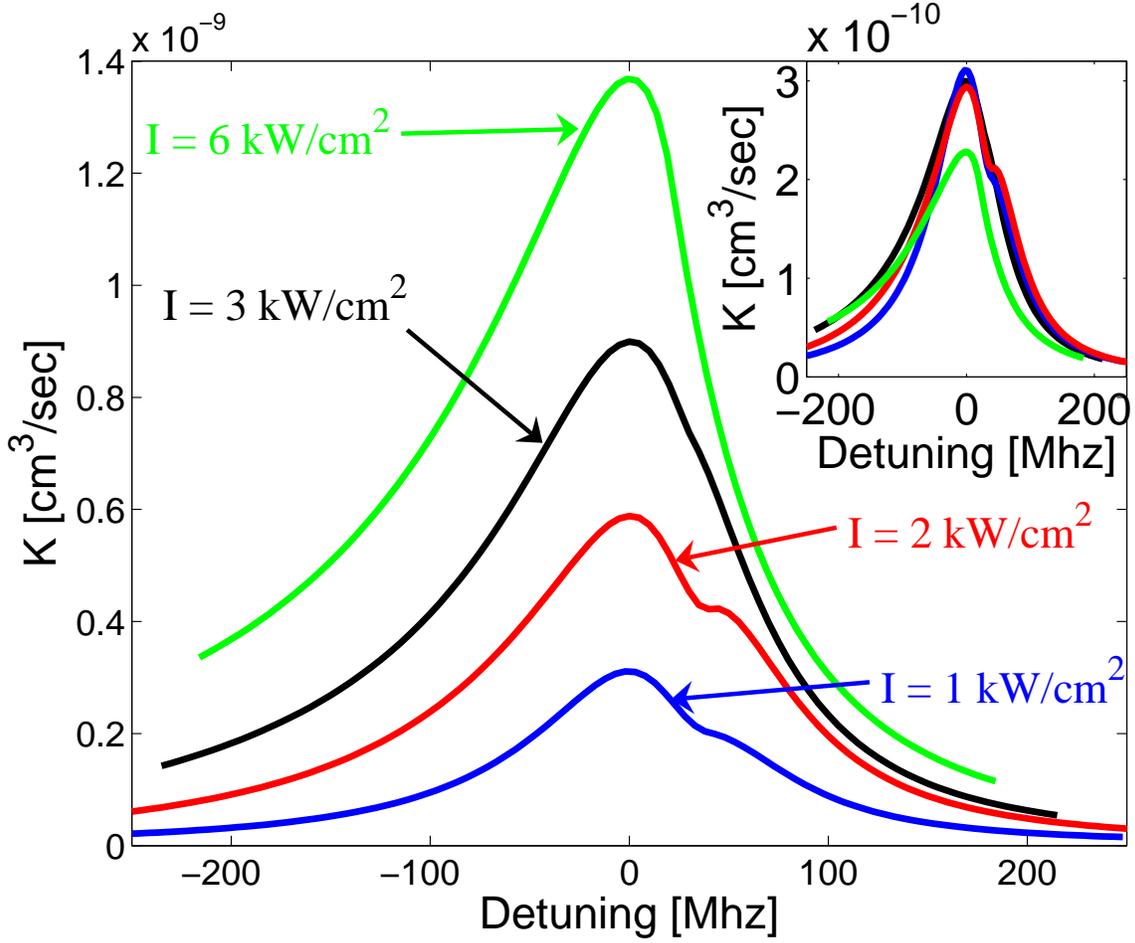, angle=0, width=\linewidth}
\caption{Lineshape around the saturation limit, using the non-perturbative method. The temperature is $T=20 \mu K$, and the blue, black, red and green lines are for intensities of 1, 2 3, and  6 $kW/cm^2$, respectively. All the lines are shifted to be centered to zero detuning. For the comparison, the line profiles are normalized to the line of $1 kW/cm^2$, i.e., the lines for $2,3$ and $6kW/cm^2$ are divided by $2,3$ and $6$.}
    \label{fig:line_high_us}
\end{figure}

\begin{figure}[htbp]
\epsfig{file=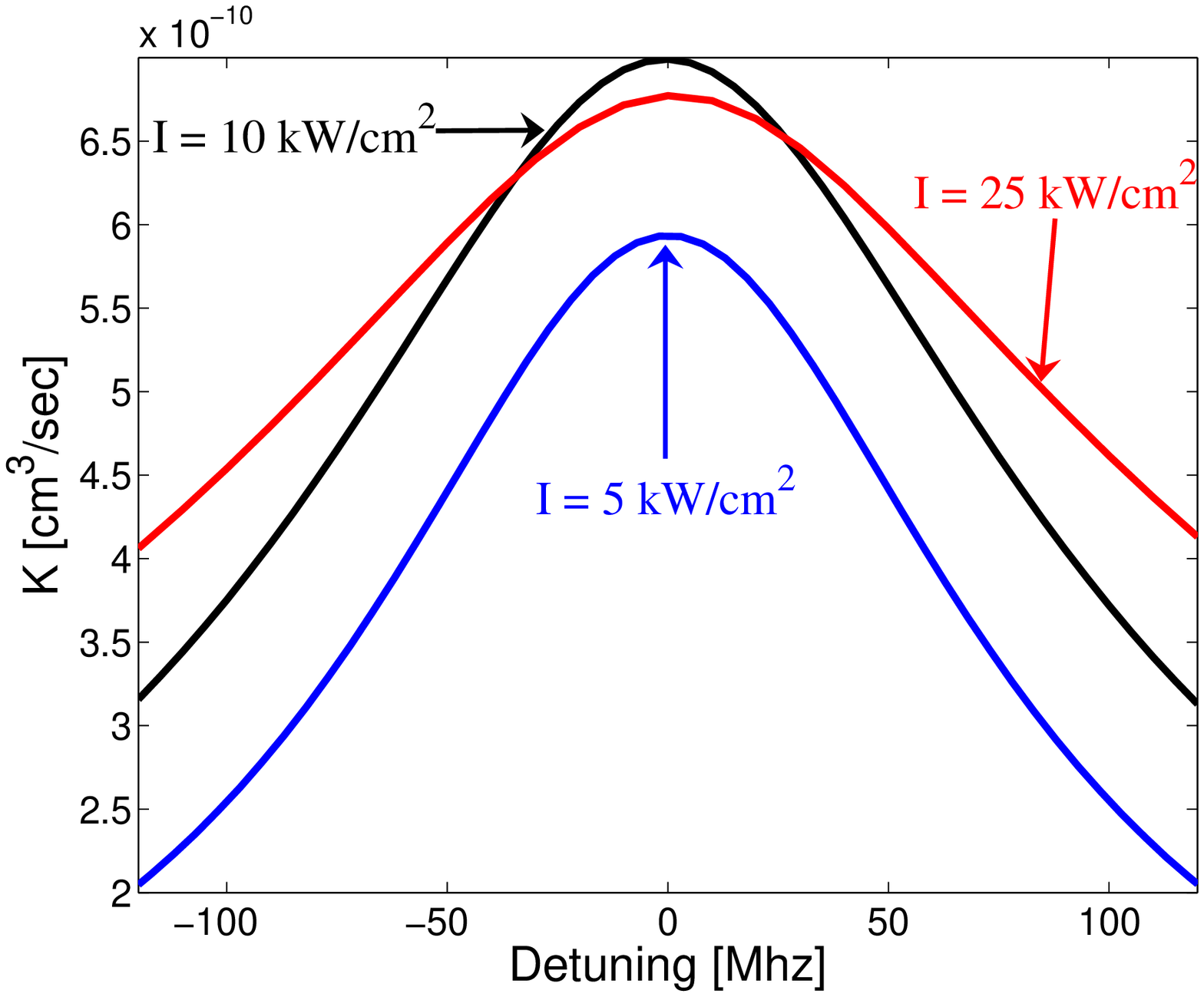, angle=0, width=\linewidth}
\caption{Same as figure \ref{fig:line_high_us}, using the perturbative method. The temperature is $T=20 \mu K$, and blue, black and red lines are for intensities of 5, 10 and  25 $kW/cm^2$, respectively. All the lines are shifted to be centered to zero detuning.}
    \label{fig:line_high_BJ}
\end{figure}

A better understanding for the origin of the inverted behavior could be acquired by an inspection of the lineshapes near the threshold. Figure \ref{fig:line_high_us} exhibits the lineshapes computed with  the non-perturbative method around the saturation regime. The four lines corresponds to the intensities of 1,2,3 and 6 kW/cm$^2$ in blue, red, black and green, respectively. All the lines are centered to zero detuning for comparison. A shoulder at the positive tail of the Lorentzian lineshape starts to develop as saturation is approached. Beyond saturation, the shoulder takes over the Lorentzian lineshape at the positive tail of the line to create an unsymmetrical lineshape. The power broadening at the negative detuning tail differs from the positive one, and the unsymmetrical form of the lineshape is more pronounced with the increase of the intensity. Figure \ref{fig:line_high_BJ} presents the lineshapes close to and beyond the saturation limit of the perturbative method, for 
 $T=20 \mu K$. The lines are centered to zero detuning. Beside the obvious power broadening, the lineshapes remain symmetrical even for high intensities.\\

As a first order perturbative method, Bohn and Julienne model takes into account only single path processes. That is, the rate constant measures the population flux from the unbound continuum levels to the bound level in the excited state. Any flux that is transfered to the bound level is lost at a rate proportional to the natural linewidth of the bound level. For low temperatures, the thermal energy width is narrow. The loss rate for the off resonance laser is determined only by the distance from the threshold. Positive and negative detuning are identical in that sense and the lineshape is therefore symmetric. For higher temperatures, the population spread is larger and the inner structure of the continuum start to play a role. This is the origin for the deviation from a Lorentzian lineshapes which presented in figure 5 of \cite{bohn99}, for temperatures of $1mK$. This non symmetric form is expected to be washed out at low temperatures and high intensities, where the thermal wi
 dth is much smaller then the radiative width.\\

In the non-perturbative method, all paths are being taken into account. At large intensity, multi-paths processes can take place. A flux could bounce  from the bound level back to the continuum and vice verse. As can be seen in figure \ref{fig:coupling_cont}, the coupling profile indicates for a maximum of the interaction around $10^{-2}$cm$^{-1}$. The shoulder of the lineshape appears at a detuning equal to the maximum. Another support for this interpretation could be found in the energy dependence of the transition probability, $P_{(g,n') \to j}$. At low intensity single path processes take place and the transition probability is significant only close to the threshold. At large intensities other parts of the profile play a role,  $P_{(g,n') \to j}$ deviates from a symmetric Lorentzian and peaks also at the energy of the maximum of the interaction profile. A further increase of the intensity enhances even further the weight of the non-linear effects on the line shape, and so
  the lineshape at positive detuning is rolled by non linear effects. A clue to support this result could be seen maybe at the apparent unsymmetrical lineshapes that were measured for the saturation limit of cesium in a MOT(see \cite{kraft2005}). 

\section{Discussion and Summary}
In the present paper we have explored the advantage of grid methods for performing a non-perturbative calculation of photoassociation lineshapes, energy shifts and rates.
Several physical properties determine the energy scales that are involved in the PA experiment and characterize the outcomes of the various observables. The energy shift, which is the value of the detuning at which a photoassociation lineshape peaks, is the parameter which is the less sensitive to the structure of the interaction profile and the thermal population. In order to calculate it accurately one has to consider an interaction energy which goes up to several hundreds of Kelvins and to include all the manifolds which participate in the interaction with the bound level. The interaction energy is independent of the thermal occupation of these levels. A full non-perturbative treatment for the energy shift is therefore extremely difficult, but fortunately, it is also  unnecessary. The advantage of the grid methods in calculating energy shifts is minor and they could be obtained efficiently and accurately by perturbative methods.\\

For low laser intensities, the natural linewidth of the bound level, which is usually at the order of hundred of MHz, or tens of milliKelvin, is the next energy scale in the hierarchy. The width of the lineshape and the scaling of the peak height are determined by this parameter at low intensities. As the intensity of the laser increases, the radiative coupling, or the Rabi frequency of the given scattering energy $\Omega(E)$ begin to dominates, and lead to power broadening and saturation.\\

\begin{table}
\begin{tabular}{|c||c||c|}
	\hline
      Notation & Parameter & Order of magnitude(Kelvins)   \\
	\hline
$E_{I}$ & Energy shifts interaction energy  & $ 10^1 - 10^3$ \\
$\gamma$ & Natural linewidth & $10^{-1}-10^{-3}$ \\
$\Omega_E$ &Stimulated emission rate & Depends on $\Delta$ and $I$ \\
$E_{T}$ & Thermal Energy & $10^{-7}-10^{-4}$ \\
	\hline 
	\end{tabular}
\caption{Parameters involve in PA and their typical energy scales in Kelvins.}
\end{table}

The thermal energy is commonly the smallest energy scale. It ranges from about a hundred of nanokelvins for the coldest condensates, to a few hundreds microkelvins, for `hot' MOTs. For low temperatures, the narrow width of the populated energy band makes the coupling profile to be almost insignificant. Saturation could then be described quite well with the perturbative methods. As the temperature increases, the inner structure of the continuum is encrypted into deviations of the lineshape from a symmetric Lorentzian, and an inverted region of $K_{max}$ as the intensity increases. Some of these deviations, especially the ones which are visible at the low intensity limit, for relatively high temperatures, could be well described by the single-path perturbative models. Other features which result only at high intensities, could be calculated only with the help of the multi-path model.\\

The non perturbative model that we presented in this paper can serve as a useful tool to model easily other cases where non trivial lineshapes appear. These include: 
\begin{itemize}
\item{Photoassociation close to threshold.}
\item{Accidentally degenerate or almost degenerate levels that originate from different electronic states. In both cases, interaction between several levels in the excited state need to be put in the framework of a higher order model. The case of resonant coupling photoassociation \cite{slava00a,dion01,kerman04,bergeman2006} is a good example.}
\item{Active interaction between multiple continuum manifolds, where more than a single continuum is populated or depopulated during the process. One possibility is when more then one partial wave participates in the scattering, e.g., the $D$ resonance in $^85$Rb (see ref. \cite{crubellier2006}). In cases of lower barriers, higher temperatures, or in the cases of processes that take place by explicitly populating the continuum, highly structured lineshapes might be resolved by our method.}
\end{itemize}

Finally, the dynamical approach which is the basis for the present study is an important building block for a future comprehensive time dependent description of the many body interaction in the description of PA. 

\subsection*{Acknowledgments} This work was supported by the Israel Science Foundation and by the
European Commission in the frame of the Cold Molecule Research Network under Contract
No. HPRN-CT-2002-00290. The Fritz Haber Center is supporter by the Minerva Gesellschaft fur
die Forschung GmbH Munchen, Germany. 

\renewcommand{\theequation}{A-\arabic{equation}}
\setcounter{equation}{0}
\bibliographystyle{jcp}

\begin{thebibliography}{10}
\newcommand{\enquote}[1]{`#1'}

\bibitem{thorsheim87}
H.~R. Thorsheim, J.~Weiner, and P.~S. Julienne.
\newblock \enquote{Laser-induced photoassociation of ultracold sodium atoms.}
\newblock Phys. Rev. Lett. {\bf 58}, 2420 (1987).

\bibitem{jones2006}
K.~M. Jones, E.~Tiesinga, P.~D. Lett, and P.~S. Julienne.
\newblock \enquote{Ultracold photoassociation spectroscopy: long range
  molecules and atomic scattering.}
\newblock Rev. Mod. Phys. {\bf 78}, 483 (2006).

\bibitem{pillet97}
P.~Pillet, A.~Crubellier, A.~Bleton, O.~Dulieu, P.~Nosbaum, I.~Mourachko, and
  F.~Masnou-Seeuws.
\newblock \enquote{Photoassociation in a gas of cold alkali atoms. {I:
  P}erturbative quantum approach.}
\newblock J. Phys. B {\bf 30}, 2801 (1997).

\bibitem{napolitano97}
R.~Napolitano.
\newblock \enquote{A two-state model for controlling scattering lengths and
  photoassociation spectral line shapes alkali-metal atoms by resonant light in
  the regime of finite ultracold temperature.}
\newblock Brazilian Journal of Physics {\bf 27}, 162 (1997).

\bibitem{cote98}
R.~C\^ot\'e and A.~Dalgarno.
\newblock \enquote{Photoassociation intensities and radiative trap loss in
  lithium.}
\newblock Phys. Rev. A {\bf 58}, 498 (1998).

\bibitem{bohn99}
J.~L. Bohn and P.~S. Julienne.
\newblock \enquote{Semianalytic theory of laser-assisted resonant cold
  collisions.}
\newblock Phys. Rev. A {\bf 60}, 414 (1999).

\bibitem{taylor2004}
E.~Taylor-Juarros, R.~C\^ot\'e, and K.~Kirby.
\newblock \enquote{Formation of ultracold polar molecules via raman
  excitation.}
\newblock Eur. Phys. J. D {\bf 31}, 213 (2004).

\bibitem{javanainen2002}
J.~Javanainen and M.~Mackie.
\newblock \enquote{Rate limit for photoassociation of a {Bose-Einstein}
  condensate.}
\newblock Phys. Rev. Lett. {\bf 88}, 090403 (2002).

\bibitem{mckenzie2002}
C.~McKenzie, J.~H. Denschlag, H.~H{\"a}ffner, A.~Browaeys, L.~E. de~Araujo,
  F.~K. Fatemi, K.~M. Jones, J.~E. Simsarian, D.~Cho, A.~Simoni, E.~Tiesinga,
  P.~S. Julienne, K.~Helmerson, P.~D. Lett, S.~L. Rolston, and W.~D. Phillips.
\newblock \enquote{Photoassociation of sodium in a {Bose-Einstein} condensate.}
\newblock Phys. Rev. Lett. {\bf 88}, 120403 (march 2002).

\bibitem{prodan2003}
I.~D. Prodan, M.~Pichler, M.~Junker, R.~J. Hulet, and J.~L. Bohn.
\newblock \enquote{Intensity dependence of photoassociation in a quantum
  degenerate atomic gas.}
\newblock Phys. Rev. Lett. {\bf 91}, 080402 (august 2003).

\bibitem{kraft2005}
S. D. Kraft, M. Mudrich, M. U. Staudt, J. Lange, O. Dulieu, R. Wester, and M. Weidemuller
\newblock \enquote{Saturation of {Cs$_2$} photoassociation in an optical dipole trap.}
\newblock Phys. Rev. A {\bf 71}, 013417 (2005).

\bibitem{naidon2006}
P.~Naidon and F.~Masnou-Seeuws.
\newblock \enquote{Photoassociation and optical Feshbach resonances in an
  atomic {Bose-Einstein} condensate: treatment of correlation effects.}
\newblock Phys. Rev. A {\bf 73}, 043611 (april 2006).

\bibitem{kohler2003b}
T.~K\"{o}hler, T.~Gasenzer, and K.~Burnett.
\newblock \enquote{Microscopic theory of atom-molecule oscillations in a
  {Bose-Einstein} condensate.}
\newblock Phys. Rev. A {\bf 67}, 013601 (2003).

\bibitem{naidon2003}
P.~Naidon and F.~Masnou-Seeuws.
\newblock \enquote{Pair dynamics in the formation of molecules in a
  {B}ose-{E}instein condensate.}
\newblock Phys. Rev. A {\bf 68}, 033612.

\bibitem{kgrid96}
R.~Kosloff.
\newblock \enquote{{Quantum Molecular Dynamics on Grids}.}
\newblock In R. E. Wyatt and J. Z. Zhang, editors, Dynamics of Molecules and Chemical Reactions, pages 185-230, Marcel Dekker, New York (1996). 
 
\bibitem{fatal}
E. ~Fattal, R. ~Baer and R. ~Kosloff,
\newblock \enquote{{Phase Space Approach for Optimizing Grid Representation: The Mapped Fourier Method}.}
\newblock Phys. Rev. E {\bf 53}, 1217 (1996).

\bibitem{slava99}
V.~Kokoouline, O.~Dulieu, R.~Kosloff, and F.~Masnou-Seeuws.
\newblock \enquote{{Mapped Fourier methods for long range molecules:
  Application to perturbations in the {Rb$_2(0_u^+$)} spectrum}.}
\newblock J. Chem. Phys. {\bf 110}, 9865 (1999).

\bibitem{willner2004}
K.~Willner, O.~Dulieu, and F.~Masnou-Seeuws.
\newblock \enquote{A mapped sine grid method for long range molecules and cold
  collisions.}
\newblock J. Chem. Phys. {\bf 120}, 548 (2004).

\bibitem{dion01}
C.~M. Dion, C.~Drag, O.~Dulieu, {Laburthe Tolra, B.}, F.~Masnou-Seeuws, and
  P.~Pillet.
\newblock \enquote{Resonant coupling in the formation of ultracold ground state
  molecules via photoassociation.}
\newblock Phys. Rev. Lett. {\bf 86}, 2253 (2001).

\bibitem{kerman04}
A.~J. Kerman, J.~M. Sage, S.~Sainis, T.~Bergeman, and D.~DeMille.
\newblock \enquote{Photoassociation and state-selective detection of ultracold
  $RbCs$ molecules.}
\newblock Phys. Rev. Lett. {\bf 92}, 153001 (2004).

\bibitem{azzizi2004}
S.~Azizi, M.~Aymar, and O.~Dulieu.
\newblock \enquote{Prospects for the formation of ultracold ground state polar
  molecules from mixed alkali atom pairs.}
\newblock Eur. Phys. J. D {\bf 31}, 195 (december 2004).

\bibitem{sage2005}
J.~M. Sage, S.~Sainis, T.~Bergeman, and D.~DeMille.
\newblock \enquote{Optical production of ultracold polar molecules.}
\newblock Phys. Rev. Lett. {\bf 94}, 203001 (2005).

\bibitem{slava00a}
V.~Kokoouline, O.~Dulieu, and F.~Masnou-Seeuws.
\newblock \enquote{Theoretical treatment of channel mixing in excited {Rb$_2$}
  and {Cs$_2$} ultra-cold molecules. Perturbations in $0_u^+$ photoassociation
  and fluorescence spectra.}
\newblock Phys. Rev. A {\bf 62}, 022504 (2000).

\bibitem{slava00b}
V.~Kokoouline, O.~Dulieu, R.~Kosloff, and F.~Masnou-Seeuws.
\newblock \enquote{Theoretical treatment of channel mixing in excited {Rb$_2$}
  and {Cs$_2$} ultra-cold molecules : determination of predissociation
  lifetimes with coordinate mapping.}
\newblock Phys. Rev. A {\bf 62}, 032716 (2000).

\bibitem{bergeman2006}
T.~Bergeman, J.~Qi, D.~Wang, Y.~Huang, H.~K. Pechkis, E.~E. Eyler, P.~L. Gould,
  W.~C. Stwalley, R.~A. Cline, J.~D. Miller, and D.~J. Heinzen.
\newblock \enquote{Photoassociation of $^{85}${Rb} atoms into 0$_u^+$ states near
  the 5{S} +5{P} atomic limit.}
\newblock J. Phys. B: At. Mol. Opt. Phys. {\bf 39}, S813 (2006).

\bibitem{lisdat01}
C.~Lisdat, O.~Dulieu, H.Knockel, and E.~Tiemann.
\newblock \enquote{Inversion analysis of {K$_2$} coupled electronic states with
  the {F}ourier grid method.}
\newblock Eur. Phys. J. D {\bf 17}, 319 (2001).

\bibitem{pellegrini2003}
P.~Pellegrini.
\newblock {\em Aspects th\'eoriques de la formation de mol\'ecules froides.
  Application \`a la photoassociation dans un condensat de Bose-Einstein\/}.
\newblock Th\'ese de doctorat (Ph.D thesis), Universit\'e Paris-Sud, Centre
  d'Orsay (2003).

\bibitem{naidon2004}
P.~Naidon.
\newblock {\em Etude th\'eorique de la formation de mol\'ecules diatomiques
  dans un condensat par photoassociation\/}.
\newblock Th\'ese de doctorat (ph.d thesis), Universit\'e Pierre et Marie Curie
  Paris VI (2004).

\bibitem{portier2006}
M.~Portier, S.~Moal, J.~Kim, M.~Leduc, C.~Cohen-Tannoudji, and O.~Dulieu.
\newblock \enquote{Analysis of light-induced frequency shifts in the
  photoassociation of ultracold metastable helium atoms.}
\newblock J. Phys. B: At. Mol. Opt. Phys. {\bf 39} S881 (2006).


\bibitem{koch06}
C. P. Koch, R. Kosloff, E. Luc-Koenig, F. Masnou-Seeuws, and A. Crubellier
\newblock \enquote{Photoassociation with chirped laser pulses : Calculation of the absolute number of molecules per pulse
.}
\newblock J. Phys. B: At. Mol. Opt. Phys. , {\bf 39} S1017 (2006).

\bibitem{nest}
M. Nest, H.-D. Meyer.
\newblock 
\newblock Chem. Phys. Lett. {\bf 352} 486 (2002).

\bibitem{cohen73}
F.~Laloe. Claude~Cohen-Tannoudji, Bernard~Diu.
\newblock {\em {\it M\'ecanique Quantique)}\/}.
\newblock (Hermann, Paris, 1973).

\bibitem{numerical_recipies} 
W.H. Press, S.A. Teukolsky, W.T. Vetterling, and B.P. Flannery.
\newblock {\em {\it Numerical Recipes)}\/}.  
\newblock (Second edition, Cambridge University Press, 1995).

\bibitem{fano61}
U.~Fano.
\newblock \enquote{Effect of configuration interaction on intensities and
  phaseshifts.}
\newblock Phys. Rev. {\bf 124}, 1866 (1961).

\bibitem{du89}
M.~L. Du, A.~Dalgarno, and M.~J. Jamieson.
\newblock \enquote{Level shifts of discrete states embedded in a continuum.}
\newblock J. Chem. Phys. {\bf 5}, 2980 (1991).

\bibitem{Bohn99}
J. ~L. Bohn and P. S. Julliene
\newblock Phys. Rev. A {\bf 60}, 414 (1999).


\bibitem{crubellier2006}
A.~Crubellier and E.~Luc-Koenig.
\newblock \enquote{Threshold effects in the photoassociation of cold atoms :
  {$R^{-6}$} model in the milne formalism.}
\newblock J. Phys. B {\bf 39}, 1417 (2006).

\bibitem{nagel2005}
S. B. Nagel, P. G. Mickelson, A. D. Saenz, Y. N. Martinez, Y. C. Chen, T. C. Killian, P. Pellegrini, and R.~C\^ot\'e
\newblock \enquote{Photoassociative Spectroscopy at Long Range in Ultracold Strontium.}
\newblock Phys. Rev. Lett. {\bf 94}, 083004 (2005).

\end{thebibliography}

\section*{APPENDIX}
Two purposes are the objects of this appendix: the numerical details of the sampling of the continuum profile ${\cal F}^e (E)$ (see eq. (\ref{eq:FC_profile})), and the demonstration of the energy scale that is needed for the calculation of the energy shift.
The sampling of the continuum profile ${\cal F}^e (E)$ could be done in principle by a global or local numerical sampling procedure (see for example \cite{numerical_recipies}).
The common procedures take advantage of an a-priori knowledge about the given function to be sampled. In this work we limited ourselves to a naive approach and determined the sampling of the profile a-posteriori according to the results that were obtained with a given basis. The criterion we employed was the numerical convergence of the integral over a given moment of the interaction profile ${\cal F}^e (E)$:
\begin{equation}
{\cal{I}} = \int\limits_{0}^{E_{cut}} { {[{\cal F}^e (E)]^n dE}} \approx \sum_j \Delta(E_j)[{\cal F}^e (E_j)]^n
\end{equation}
where $n$ is the order of the moment, $E_j$ is the discrete sampling energy points, and $\Delta(E_j)$ is the energy difference between two adjacent points. The upper bound for the profile $E_{cut}$ is discussed below. The procedure initiates with a small set of points, and a new point is added or rejected between any two old points according to its influence on the value of ${\cal{I}}$. The quality of the procedure is determined by the tolerance parameter $\tau = \left| {\cal{I}}^{new}/{\cal{I}}^{old}-1 \right|$.
The number of basis functions for the sampling the $S$ continuum, as a function of the tolerance parameter is presented in figure A1. The various lines present different moments. \\

\begin{figure}[htbp]
\epsfig{file=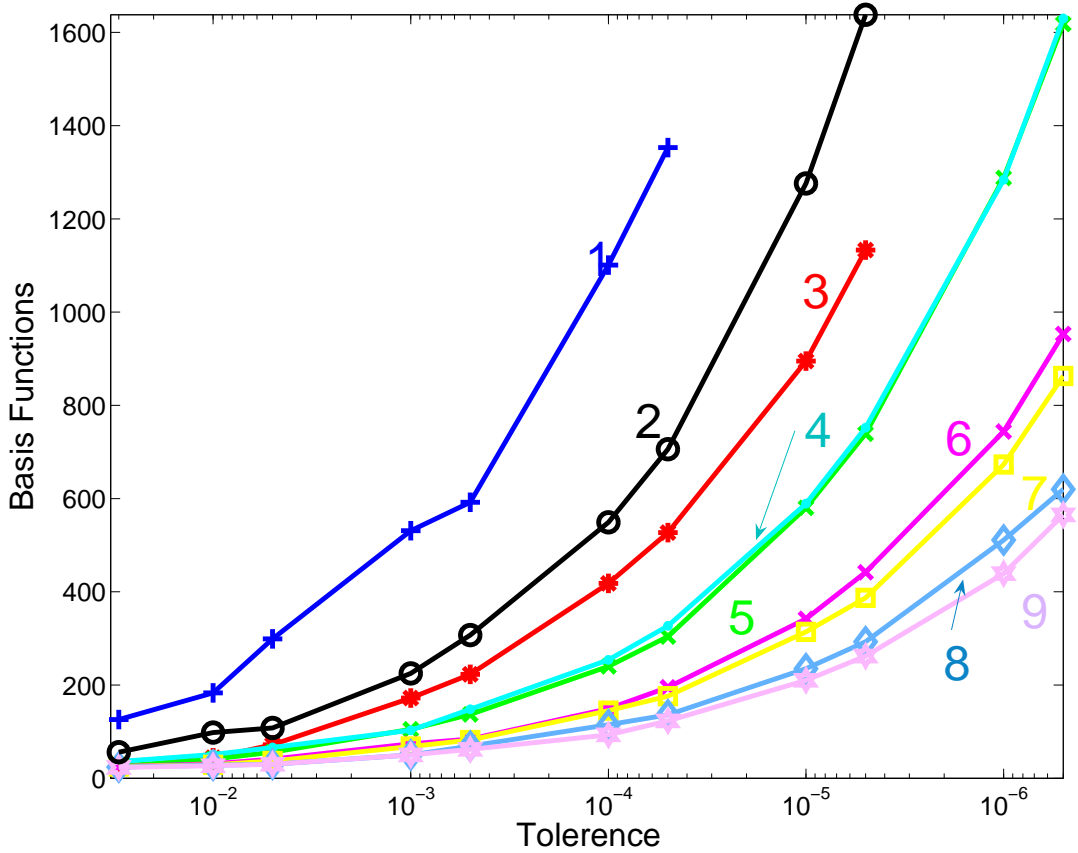, angle=-90, width=\linewidth}
\caption{Number of basis function vs. the tolerance parameter $\tau$. The numbers near the lines stand for different moments of the interaction profile ${\cal F}^e (E)$. The box normalized from which the original profile is deduced contains 920 basis functions.}
    \label{fig:app1}
\end{figure}

As can be understood, the convergence of the sampling becomes more efficient for higher moments, which emphasize more the features in the profile. Too high moments, however, smear completely the functional structure and so an optimum moment have to be found. We have to comment here that in order to increase the resolution of the basis set with respect to the thermally populated energy near the threshold (for $E < Nk_{B}T$, N is usually 10), we inserted a small energy grid (typically 50-100 basis functions) with constant energy step at close-to-threshold energies.\\

In figure A2 the resulted energy shifts are plotted as a function of the tolerance parameter for the various moment orders. The numerically exact value which was calculated using the perturbative method is $27.8 MHz$. The values for moment order above the $5^{th}$, convergence to this values within less then 5\% for tolerance value of $5\times 10^{-5}$. The basis set that was taken in this paper uses $\tau = 5\times 10^{-7}$ and $n = 9$. Larger moments than the chosen exhibit slower convergence and demand much smaller values of $\tau$ to produce an adequate basis set. \\

\begin{figure}[htbp]
\epsfig{file=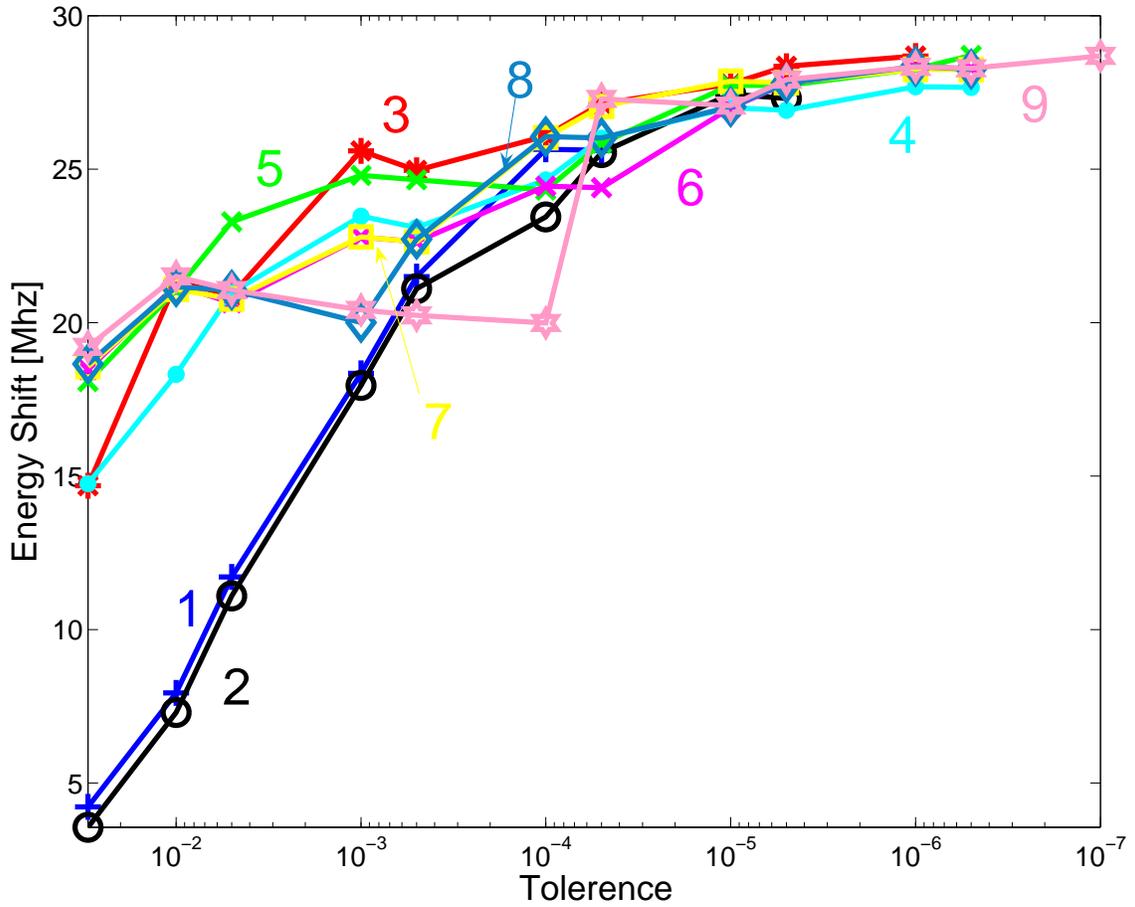, angle=-90, width=\linewidth}
\caption{Calculated energy shift vs. the tolerance parameter $\tau$. The intensity is $100W/cm^2$, the temperature is $2 \mu K$. Only calculation with respect to the $S$ continuum are presented. The continuum is sampled up to energy of $300K$. The numbers near the lines stand for different moments of the interaction profile ${\cal F}^e (E)$. }
    \label{fig:app2}
\end{figure}

We must remark also, that the only physical parameter which demands such a care in the sampling is the energy shift. The values for the lifetime are much less influenced by the size of the basis, and converged even at $\tau$ as small as $1\times10^{-3}$. To demonstrate the importance of very large energy scale that are needed to calculate the energy shift correctly we present in figure A3 the calculated energy shift as a function of $E_{cut}$ the upper bound of the energy grid. Convergence of the energy shift is achieved only when reaching hundreds of Kelvins. Note, again, that only the energy shift exhibit a sensitivity for such high energies. For low laser intensities the lifetime is found to converge $E_{cut}$ at the order of the thermal energy grid. Convergence near the saturation limit demands extension of the grid up to the first minimum of the coupling profile, i.e.,$10^{-2}cm^{-1}$, or tens of milikelvins. A basis set for such energies contained less then 150 function
  and is adequate for calculating most of the lineshapes features, except for energy shifts.\\

\begin{figure}[htbp]
\epsfig{file=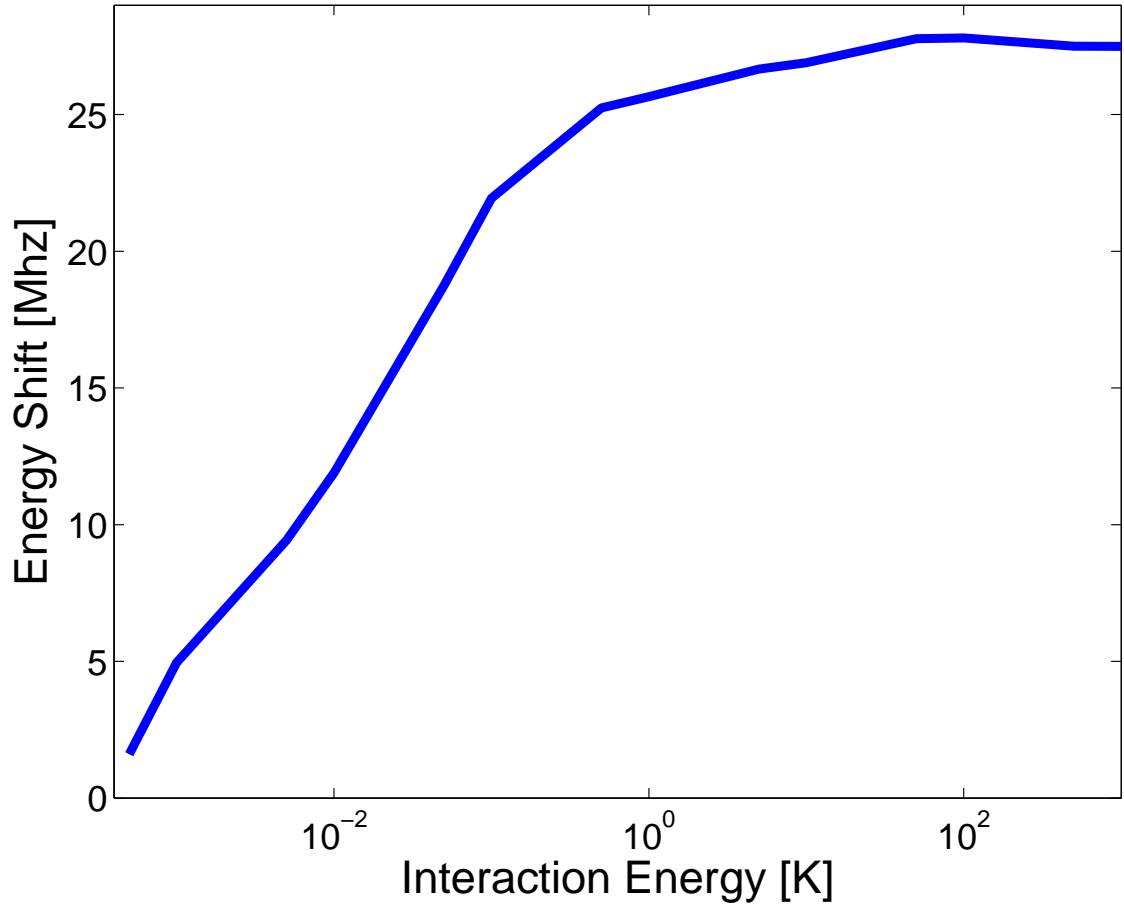, angle=0, width=\linewidth}
\caption{Calculated energy shift vs. highest continuum energy to sample. The parameters are identical to figure A2. The ninth moment with tolerance of $5\times 10^{-7}$ is taken.}
    \label{fig:app3}
\end{figure}

\end{document}